\PassOptionsToPackage{top=25.4mm, bottom=25.4mm, left=25.4mm, right=25.4mm}{geometry}

\documentclass[referee,sn-nature]{sn-jnl} 

\usepackage{graphicx}
\usepackage{multirow}
\usepackage{amsmath,amssymb,amsfonts}
\usepackage{amsthm}
\usepackage{mathrsfs}
\usepackage[title]{appendix}
\usepackage{xcolor}
\usepackage{textcomp}
\usepackage{manyfoot}
\usepackage{booktabs}
\usepackage{hyperref}
\usepackage{algorithm}
\usepackage{algorithmicx}
\usepackage{algpseudocode}
\usepackage{listings}
\usepackage{lineno}
\usepackage{siunitx}
\theoremstyle{thmstyleone}

\theoremstyle{thmstyletwo}

\theoremstyle{thmstylethree}

\usepackage{etoolbox}

\makeatletter
\newcommand{\preabstractnote}{}
\let\SN@orig@printabstract\printabstract
\renewcommand{\printabstract}{%
  \ifx\preabstractnote\@empty\else
    \par\addvspace{6pt}{\small \noindent\preabstractnote\par}\addvspace{12pt}%
  \fi
  \SN@orig@printabstract
}
\makeatother

\begin{document}

\title[Article Title]{Electron-beam-induced Contactless Manipulation of Interlayer Twist in van der Waals Heterostructures}

\author*[1,2,3]{\fnm{Nicola} \sur{Curreli}}

\author[2]{\fnm{Tero S.} \sur{Kulmala}}

\author[2,4]{\fnm{Riya} \sur{Sebait}}

\author[1,3]{\fnm{Nicolò} \sur{Petrini}}

\author[5]{\fnm{Matteo Bruno} \sur{Lodi}}

\author[2]{\fnm{Roman} \sur{Furrer}}

\author[5]{\fnm{Alessandro} \sur{Fanti}}

\author[2,6]{\fnm{Michel} \sur{Calame}}

\author[1,3]{\fnm{Ilka} \sur{Kriegel}}

\affil*[1]{\orgdiv{Department of Applied Science and Technology}, \orgname{Polytechnic University of Turin}, \orgaddress{\street{Corso Duca degli Abruzzi 34}, \city{Turin}, \postcode{10129}, \country{Italy}}}

\affil[2]{\orgdiv{Transport at Nanoscale Interfaces Laboratory}, \orgname{Empa, Swiss Federal Laboratories for Materials Science and Technology}, \orgaddress{\street{Ueberlandstrasse 129}, \city{Dübendorf}, \postcode{8600}, \country{Switzerland}}}

\affil[3]{\orgdiv{Functional Nanosystems}, \orgname{Italian Institute of Technology}, \orgaddress{\street{Via Morego 30}, \city{Genoa}, \postcode{16163}, \country{Italy}}}

\affil[4]{\orgdiv{Nanotechnology at Surfaces Laboratory}, \orgname{Empa, Swiss Federal Laboratories for Materials Science and Technology}, \orgaddress{\street{Ueberlandstrasse 129}, \city{Dübendorf}, \postcode{8600}, \country{Switzerland}}}

\affil[5]{\orgdiv{Department of Electrical and Electronic Engineering}, \orgname{University of Cagliari}, \orgaddress{\street{Via Marengo 3}, \city{Cagliari}, \postcode{09123}, \country{Italy}}}

\affil[6]{\orgdiv{Department of Physics and Swiss Nanoscience Institute}, \orgname{University of Basel}, \orgaddress{\street{Klingelbergstrasse 82}, \city{Basel}, \postcode{4056}, \country{Switzerland}}}

\abstract{
The ability to dynamically control the relative orientation of layers in two dimensional (2D) van der Waals (vdW) heterostructures represents a critical step toward the realization of reconfigurable nanoscale devices. Existing actuation methods often rely on mechanical contact, complex architectures, or extreme operating conditions, which limit their applicability and scalability. In this work, we present a proof-of-concept demonstration of contactless electrostatic actuation based on electron-beam-induced charge injection. By locally charging an insulating hexagonal boron nitride (hBN) flake on an electrically grounded graphene layer, we create an interfacial electric field that generates in-plane electrostatic torque and induces angular displacement. We validate the induced rotation through in-situ scanning electron microscopy (SEM) and twist-dependent Raman spectroscopy.
}

\keywords{van der Waals heterostructures, twist angle engineering, 2D materials, electrostatic torque, scanning electron microscopy, Raman spectroscopy, moiré superlattices, charge injection, non-contact actuation}

\renewcommand{\preabstractnote}{%
  {\small \textbf{*Corresponding Author:} Nicola Curreli (\texttt{nicola.curreli@polito.it})\par}
}

\maketitle

\section*{Introduction}
The growing field of reconfigurable nano-devices has attracted significant attention due to its potential to impact various technological areas, including advanced sensors, actuators, and next-generation electronic components \cite{lemme2020nanoelectromechanical,pham20222d}. This interest originates from the distinct physical and chemical properties of materials at the nanoscale, which often differ significantly from their bulk counterparts, allowing for new functions \cite{cao2018unconventional,chiu2014spectroscopic,shi2018interlayer,vargas2017tunable,zhang2017interlayer,cao2018correlated}. The ability to manipulate and control the state of nanoscale systems through actuation mechanisms is important to fully exploit their functionality and improve the performance of nanoscale devices \cite{lemme2020nanoelectromechanical}. A technological platform that puts these actuation principles into practice is provided by nano- and micro-electromechanical systems (N/MEMS), which integrate movable mechanical structures with their driving and sensing electronics on the same chip \cite{lemme2020nanoelectromechanical}. Among the different materials investigated within N/MEMS architectures, two-dimensional (2D) van der Waals (vdW) materials have emerged as particularly promising candidates \cite{lemme2020nanoelectromechanical}. Moreover, combining distinct 2D materials into heterostructure architectures not only allows precise tailoring of the properties of each individual component, but, more importantly, also opens pathways to unique device functionalities that would otherwise be impossible \cite{hanbicki2018double,rivera2015observation,vargas2017tunable,yankowitz2018dynamic}. 
Recent research has explored several actuation mechanisms for 2D materials, including piezoelectric \cite{ren20252d, ares2020piezoelectricity}, piezoresistive \cite{fan2019suspended}, opto-thermal \cite{steeneken2021dynamics, ryu2020microheater}, and electrostatic \cite{lemme2020nanoelectromechanical} approaches, which are frequently implemented within N/MEMS platforms to exert controlled forces or strain.
For instance, piezoelectric actuation has been explored for monolayer hexagonal boron nitride (hBN), which exhibits piezoelectric properties due to the lack of a center of symmetry, suggesting its potential for electromechanical devices \cite{ares2020piezoelectricity}. However, the piezoelectric response in ultrathin layers can be limited, and factors such as stress induced during fabrication and potential contamination can affect device performance \cite{ares2020piezoelectricity}. 
On the other hand, opto-thermal actuation presents a contactless approach to manipulating 2D materials, typically relying on the use of lasers to induce localized heating and thermal expansion \cite{steeneken2021dynamics}. For example, micro‑heaters can achieve controlled biaxial strain in 2D materials via thermal expansion \cite{ryu2020microheater}.
Nevertheless, this method can be constrained by requirements for high-temperature processing, the risk of material degradation due to heating, and the complexity of experimental setups.
Another example of actuation exploits the piezoresistive effect of suspended graphene membranes with attached silicon proof masses to realize a piezoresistive N/MEMS accelerometers \cite{fan2019suspended}. While offering mechanical control, this approach can introduce stress during fabrication, potential contamination from the substrate, and complexities in device integration.
In this context, electrostatic actuation, which relies on Coulombic forces, has been extensively explored as a versatile technique for manipulating N/MEMS due to its inherent potential for achieving high operating frequencies and its relative ease of integration into device architectures \cite{lemme2020nanoelectromechanical}. 
In fact, at the microscale, Coulomb forces arising from net charges or induced polarization can overcome the weak vdW interactions between layers \cite{gauthier2013analysis}. 
In principle, these forces can address several mechanical degrees of freedom in stacked 2D crystals, such as out-of-plane separation, in-plane strain, and relative rotation \cite{dai2020mechanics}.
The latter, quantified by the interlayer twist angle, controls the formation of moiré superlattices and thereby tunes a plethora of correlated electronic and optical phenomena \cite{cao2018unconventional, trovatello2021optical}. Deterministic twist control is therefore a primary goal; however, achieving precise and reproducible tuning of the twist angle has proven remarkably challenging. 
Indeed, most electrostatic schemes reported to date induce lateral sliding or bending rather than deterministic rotation \cite{lemme2020nanoelectromechanical}. Closing this gap is a key prerequisite for truly reconfigurable 2D heterostructures. A promising route is to tailor the charge distribution so that the resulting lateral electric field exerts an electrostatic torque, rotating the adjacent layers. Several strategies have been proposed to exploit this concept.
Solution-based methods use surface charges in liquid environments (e.g., through solvents, pH, or ionic screening) to assist in the assembly and alignment of 2D materials \cite{hill2024layered}. Charged 2D flakes in liquid can self-assemble with a tendency toward commensurate alignment due to electrostatic interactions overcoming random vdW attachment, although they offer limited post-deposition control \cite{hill2024layered}. 
Tip-based charge injection techniques provide a tool for achieving highly localized electrostatic control over 2D materials at the nanoscale \cite{vasic2012atomic}. A sufficiently high bias voltage between the conductive tip and the sample can inject charges into the 2D material by overcoming local energy barriers, primarily through electron tunneling across the thin insulating layer between the tip and the 2D material \cite{batool2023electrical}.
These techniques suffer from the significant drawback that the mechanical contact often required between the tip and the sample can introduce unwanted strain, contamination, or even physical damage to the atomically thin 2D layers. 

To overcome the limitations of the aforementioned methods, this work introduces a proof-of-concept demonstration of contactless electrostatic actuation of 2D vdW heterostructures exploiting direct charge injection via an electron beam. 
The core mechanism of our approach can be described through a nanoscale parallel-plate capacitor model. The bottom layer of the heterostructure, consisting of graphene, is electrically grounded and serves as a stator, establishing a fixed reference potential. The top layer, composed of hBN, adheres to graphene only through vdW forces, and due to the incommensurate alignment in our system, exhibits ultralow interlayer shear strength \cite{mandelli2017sliding}. Hence, for the in-plane rotational degree of freedom, the flake behaves as a mechanically decoupled rotor. 
When a focused scanning electron microscope (SEM) beam is directed onto the hBN flake, electrons accumulate on its surface, inducing a potential difference with respect to the grounded graphene. This generates an electrostatic field with both vertical and lateral components at the interface. While the vertical component drives interlayer attraction, the lateral field components are responsible for producing an in-plane torque that acts to align the hBN flake with the underlying graphene \cite{hennighausen2019probing}. The rotational response of the system proceeds until a mechanical equilibrium is achieved, defined by the balance between electrostatic torque and the mechanically resistive torque due to vdW static friction \cite{woods2016macroscopic}. 
To validate the electrostatically driven rotation, we combine in-situ SEM imaging and Raman spectroscopy as complementary non-contact diagnostics \cite{eckmann2013raman}. SEM imaging tracks the twist angle between the graphene and hBN layers in real time by directly resolving the angle between adjacent flakes. Raman spectroscopy, instead, detects the moiré super-lattice through twist-dependent shifts and splittings of the Raman peaks; the period of the moiré pattern extracted from these spectral features yields an independent quantitative estimate of the twist angle \cite{eckmann2013raman}. Agreement between the two measurements provides a robust cross-check of the contact-free actuation. 

\section*{Discussion}\label{sec2}

To experimentally demonstrate electron-beam-driven twist control, we carried out a series of systematic experiments on graphene/hBN heterostructures, where twist modulation was induced by electron beam exposure. The experimental setup consisted of a graphene stator, electrically grounded using an Imina Technologies nanoprobing platform, and a hexagonal boron nitride rotor. 
A thin gold pad was patterned on the hBN flake to serve as a charge-collection electrode, providing an equipotential surface for the accumulation of electrons delivered by the SEM beam, operated under controlled parameters (see \nameref{sec11} for details).

\begin{figure}
    \centering
    \includegraphics[width=0.7\linewidth, trim=0 275 300 0, clip]{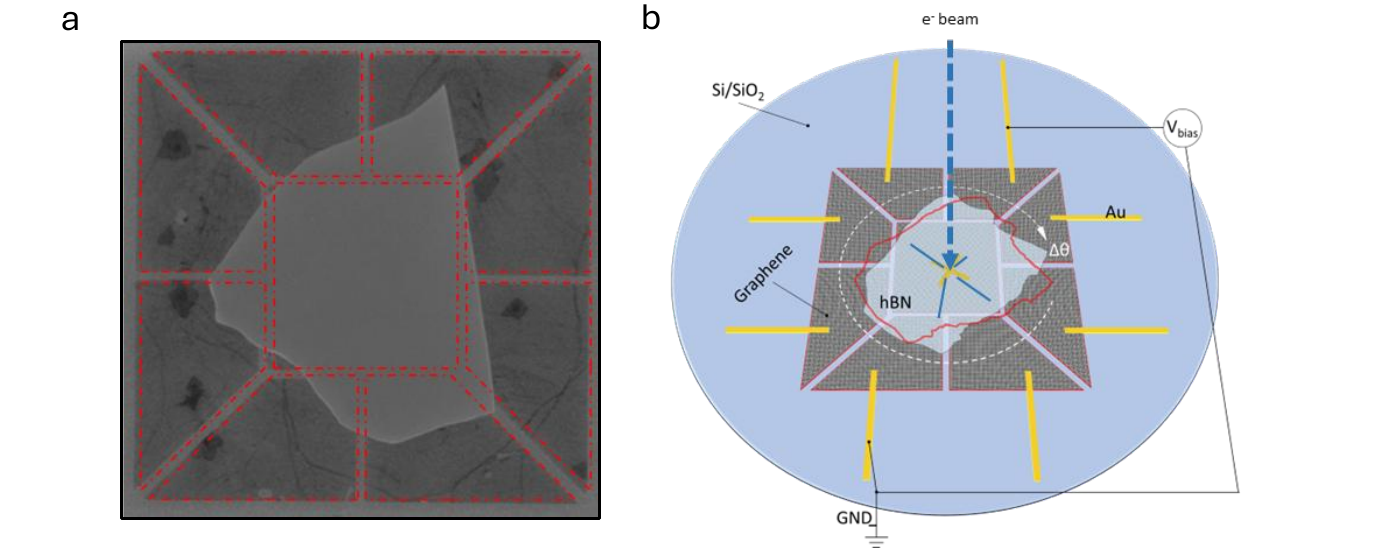}
    \caption{
a) Scanning electron micrograph of a representative graphene/hBN heterostructure. The bottom graphene layer (stator) is patterned into a square geometry. A mechanically decoupled hBN flake (rotor) is placed atop the patterned graphene (stator). The red dashed lines highlight the etched geometry of the stator.
b) Conceptual representation of the device operation under electron beam irradiation. A focused electron beam (e$^-$ beam) locally charges the hBN layer, inducing a potential difference with respect to the grounded graphene. The resulting electrostatic field at the interface comprises vertical and lateral components. The lateral field exerts an in-plane torque on the hBN rotor (curved white arrow), starting rotational motion toward a new equilibrium twist angle.}
    \label{fig:fig0}
\end{figure}

\begin{figure}
    \centering
    \includegraphics[width=\linewidth, trim=0 26 20 0, clip]{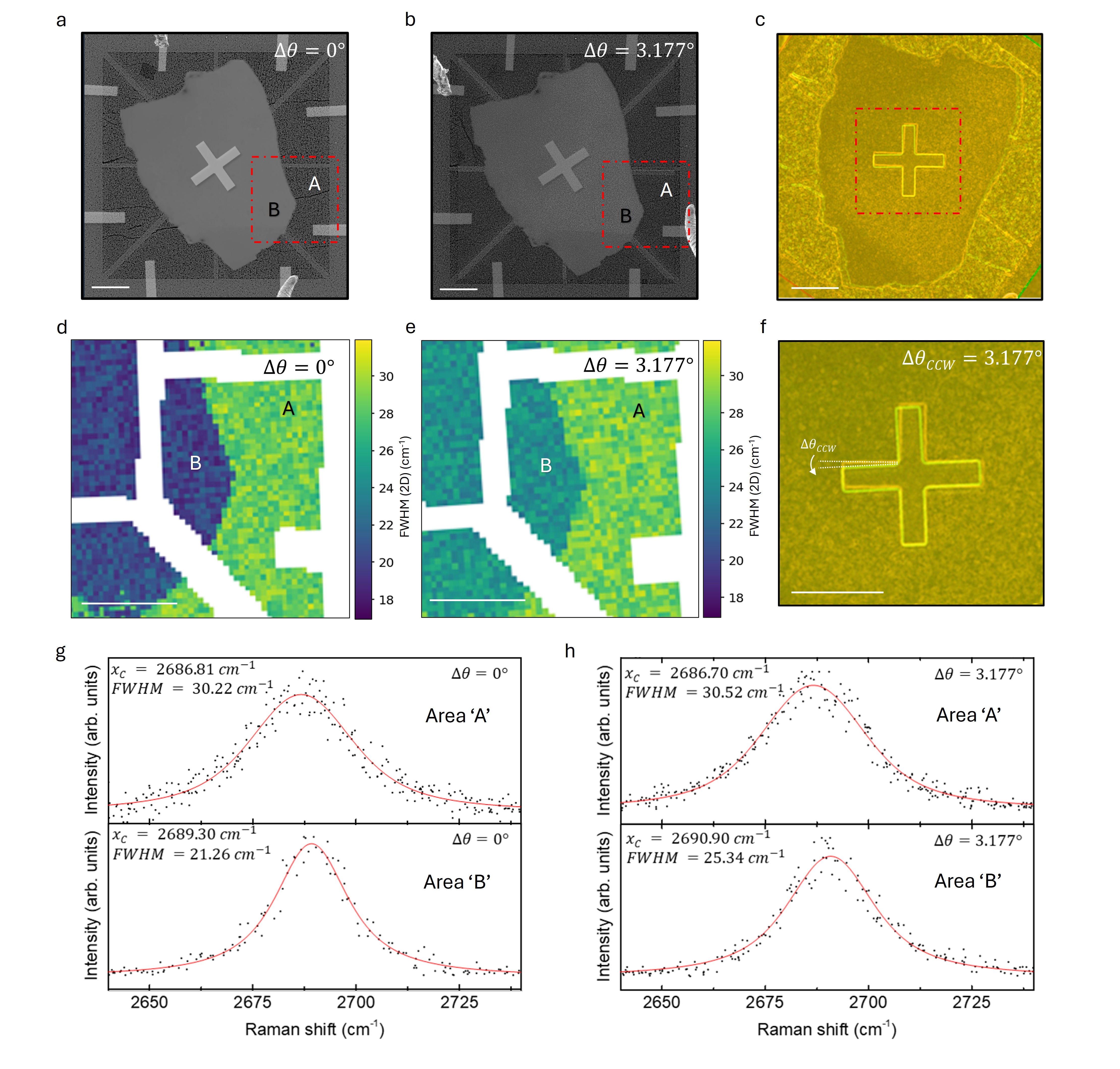}
\caption{
a) Scanning electron microscopy image of Sample S1 acquired \emph{before} electron-beam exposure; the hBN flake rests on patterned graphene pads and the red dashed rectangle marks the area analysed by Raman mapping. 
b) Scanning electron microscopy image of the same region \emph{after} electron injection, showing a counter-clockwise rotation of the flake. 
c) Overlay of images a) and b) reveals a net twist of $\Delta\theta \approx 3.18^{\circ}$. 
d) Map of the full width at half maximum (FWHM) of the graphene 2D Raman band recorded \emph{before} actuation. 
e) Corresponding FWHM map \emph{after} actuation; regions A (bare graphene) and B (graphene covered by hBN) are visible in both maps. 
f) Enlarged view from c) highlighting the measured rotation. 
g) Representative Raman spectra extracted from d). 
h) Corresponding spectra from e). 
Scale bar: 10 µm (applies to all panels).}
    \label{fig:S1}
\end{figure}

\begin{figure}
    \centering
    \includegraphics[width=\linewidth, trim=0 26 20 0, clip]{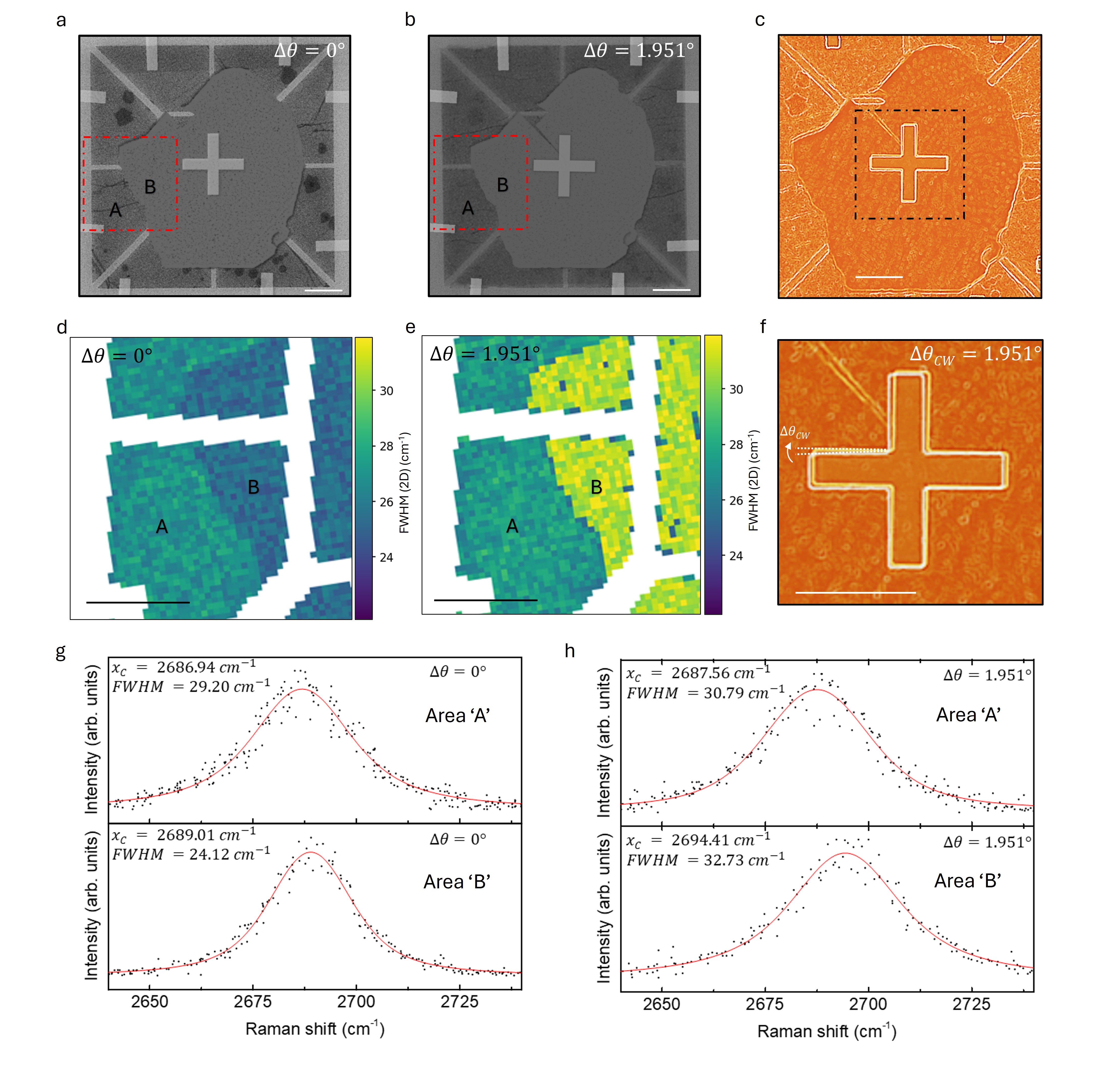}
    \caption{
a) Scanning electron microscopy image of Sample S2 acquired \emph{before} electron-beam exposure; the hBN flake rests on patterned graphene pads and the red dashed rectangle marks the area analysed by Raman mapping. 
b) Scanning electron microscopy image of the same region \emph{after} electron injection, showing a clockwise rotation of the flake. 
c) Overlay of images a) and b) reveals a net twist of $\Delta\theta \approx 1.95^{\circ}$. 
d) Map of the full width at half maximum (FWHM) of the graphene 2D Raman band recorded \emph{before} actuation. 
e) Corresponding FWHM map \emph{after} actuation; regions A (bare graphene) and B (graphene covered by hBN) are visible in both maps. 
f) Enlarged view from c) highlighting the measured rotation. 
g) Representative Raman spectra extracted from d). 
h) Corresponding spectra from e). 
Scale bar: 10 µm (applies to all panels).}
    \label{fig:S2}
\end{figure}

Figure \ref{fig:fig0} summarizes both the physical layout and the electrostatic operation principle. The final device layout is shown in Figure \ref{fig:fig0}a, which presents a representative SEM image of the graphene/hBN stack with the etched stator geometry (red dashed lines).
This specific geometry was chosen to maximise the graphene area while maintaining electrical isolation between the individually biased pads, and at the same time to ensure that the hBN flake lies entirely on graphene, avoiding any contact with the underlying Si/SiO$_2$ substrate, thus preventing pinning or strain. 
A schematic illustration of the actuation mechanism is provided in Figure \ref{fig:fig0}b, highlighting how local charge injection via electron beam leads to the generation of electrostatic torque driving the rotation of the hBN rotor. Two opposite graphene pads of the stator are biased at \( \pm V_\text{bias} \) ($<$ 1 V) while the remaining pads are grounded, establishing a lateral potential gradient across the interface. The injected charge polarizes in response to \( V_\text{bias} \), giving rise to an in-plane electrostatic field \( \mathbf{E}_\parallel \) that exerts a torque on the hBN rotor and drives it towards the equilibrium twist angle controlled by the bias polarity.
The graphene/hBN heterostructures were fabricated via established dry-transfer techniques onto Si/SiO$_2$ substrates. 
The fabrication of the heterostructures followed a multi-step approach to ensure precise patterning and stacking of the 2D materials (see Supplementary Information and Supplementary Figure S1). Monolayer graphene was synthesized on a Cu foil substrate via chemical vapor deposition (CVD) and then transferred onto a Si/SiO$_2$ substrate (extended details are reported in \nameref{sec11} and Supplementary Information). 
A square-shaped geometry (see Figure \ref{fig:fig0}) was adopted for the graphene stator, and the CVD graphene was patterned using a combination of electron beam lithography (EBL) and reactive ion etching (RIE).
Identical patterning was carried out on three separate Si/SiO$_2$ chips, each hosting a continuous monolayer graphene film. On every chip, 64 square graphene regions (50 $\times$ 50 $\mu$m$^2$) were defined, yielding a total of 192 graphene stators. From these, only the highest-quality graphene regions, free of defects such as tears, multilayer inclusions, polymer residues, or irregular etching profiles, were selected for heterostructure assembly.
Subsequently, hBN flakes ($\sim$30 nm thick, see Supplementary information for details) were exfoliated and transferred onto the selected graphene regions to implement graphene/hBN heterostructures, and gold electrodes were defined to provide independent electrical access. During transfer and electrode fabrication, several heterostructures were lost due to mechanical failure or processing issues. Ultimately, six stacks exhibited twist dynamics under SEM irradiation and were analyzed in detail (extended data in Supplementary Figures S5–S10).
The SEM was equipped with a nanoprobing system to enable both charge injection and real-time observation of interfacial dynamics within the graphene/hBN heterostructures. 
The electron beam parameters were optimized to deliver controlled energy input while minimizing the risk of beam-induced damage. Specifically, the accelerating voltage was set to 5 keV to restrict the electron penetration depth and reduce the total dose imparted to the system. The beam was defined by a 20 $\mu m$ aperture. Beam scanning was performed over a 50 $\times$ 50 $\mu m^2$ area with a dwell time corresponding to a scan rate of $\sim$1 s per frame, and the beam current was maintained below 100 pA.
From an energy-balance estimation, the current and voltage applied yield an energy density of $\sim$10$^2$ -- 10$^3$ $J m^{-2}$ per scan, corresponding to a dose of 2 -- 4 $\mu $C cm$^{-2}$ (see Supplementary Note 2 for the detailed energy dose estimation). Under these conditions, we did not observe evidence of beam-induced lattice damage, as confirmed by Raman measurements before and after exposure (see Supplementary Note 4 and Supplementary Figures S2–S4). Moreover, although the doses that cause damage are material dependent, the energy density applied in our experiments is orders of magnitude lower than values reported to induce significant morphological modifications in other 2D heterostructures, e.g., Bi$_2$Se$_3$/MoS$_2$ (3$\times$10$^6$ $J m^{-2}$ \cite{hennighausen2019probing}). This supports that the damage threshold was not reached in our heterostructures; however, cross-material comparisons should be interpreted with caution.
A fundamental aspect of the experimental design was the implementation of a well-defined electrostatic boundary condition through the electrical grounding of the graphene layer, which served as the static electrode (stator) of the heterostructure. The ground connection established a reference potential for generating a stable electric field between the graphene and the hBN layer. In the absence of this grounding condition, charge accumulation on both layers would render the capacitive interaction uncontrollable. Under electron beam irradiation, the hBN layer acted as a mechanically decoupled rotor tuned via charge injection. The actuation mechanism, based on lateral electrostatic torque generated by asymmetric charge distribution, is depicted in Figure \ref{fig:fig0}b.
Under the abovementioned conditions, SEM imaging revealed clear signs of twist modulation in our graphene/hBN stacks. 
To systematically investigate the effect of electron-beam-induced actuation on interlayer twist, we tested the selected graphene/hBN heterostructures under conditions designed to favor either clockwise (CW) and counterclockwise (CCW) rotation. 
Images acquired before and after the electrostatically induced rotation showed interlayer motion progressing toward the grounded graphene layer (the stator), consistent with electrostatically driven torque.
In representative samples S1 and S2, shown in Figure \ref{fig:S1} and Figure \ref{fig:S2}, respectively, a clear angular displacement of the hBN flake relative to the underlying graphene is observed. 

Specifically, in Sample S1, the pre- and post-rotation configurations are shown in Figure \ref{fig:S1}a and \ref{fig:S1}b, respectively (see Supplementary Figure S5 for additional characterization of Sample S1). The corresponding twist angle is quantitatively extracted in Figure \ref{fig:S1}c using a dedicated Python-based image registration procedure: the two SEM images are aligned based on a central region of interest centered on the gold cross atop the hBN flake, and a brute-force rigid rotation is applied to the post-rotation frame. The angle that minimizes the pixel-wise mean squared error between the two frames is identified as the actual twist, yielding a counterclockwise rotation of approximately 3° (extended details are reported in \nameref{sec11} and Supplementary Information). Figure \ref{fig:S1}f provides a zoomed overlay of the aligned images. 
This demonstrates that even a relatively low-dose electron injection, combined with a grounded graphene stator, can generate sufficient electrostatic torque to overcome interlayer friction, thus driving the hBN rotor into a new angular configuration.
Given its wide bandgap ($\sim$6 eV \cite{kirchhoff2022electronic}) and low intrinsic conductivity \cite{cheng2024enhanced}, hBN promotes spatially localized charge accumulation, enhanced by the limited out-of-plane transport in hBN, allowing the buildup of a quasi-static negative charge distribution. This non-uniform surface potential, in the presence of the grounded graphene stator beneath, leads to the formation of an effective electric field with both planar and perpendicular components across the heterostructure.
As a result, the electron beam irradiation leads to an electrostatic force that can overcome static friction at the interface, allowing the hBN layer to rotate into an alignment that minimizes the overall system energy. The electric field exerts an electrostatic torque that rotates the hBN layer in-plane, modulating the twist angle of the graphene/hBN superlattice until the torque is balanced by interlayer vdW friction, at which point a new equilibrium is reached \cite{woods2016macroscopic,hennighausen2019probing}. In this way, the electron beam acts as a remote control for manipulating 2D interfaces.
In addition, although SEM imaging allowed us to clearly track the incremental rotation of the hBN rotor relative to the graphene stator ($\Delta \theta$), it did not provide direct access to the crystallographic twist angle ($\theta$) that defines the moiré superlattice between the two layers.
This is primarily due to the lack of crystallographic reference for the individual layers before to beam exposure.
This limitation prevented direct measurement of the moiré angle. However, indirect estimation of the moiré periodicity was achieved via spatially resolved Raman spectroscopy, by exploiting frequency shifts and peak broadening in the graphene 2D modes. In fact, these spectroscopic signatures are sensitive to the presence of moiré-induced modulation of the phonon dispersion and allowed us to estimate the moiré wavelength, and hence infer a range of possible twist angles, even in the absence of direct lattice imaging.
Hence, to complement and further validate the results obtained from electron-beam-induced twist experiments, we performed Raman spectroscopy maps comparing two distinct regions of the sample: one consisting of monolayer graphene on the substrate, and another where the same graphene layer was covered by an hBN flake, forming a graphene/hBN heterostructure.
In graphene/hBN heterostructures, the moiré pattern generated by lattice mismatch or a relative rotation angle $\theta$ induces a long-wavelength periodic potential. The associated moiré wavelength $\lambda_M$ modulates the electronic structure and introduces inhomogeneous strain in graphene, which is reflected in the Raman response through broadening and asymmetry of the 2D peak. These spectral features have been correlated with both $\lambda_M$ and $\theta$ \cite{ferrari2013raman,eckmann2013raman,cheng2015raman,finney2019tunable}. Notably, Eckmann et al. \cite{eckmann2013raman} reported a linear relationship between the FWHM of the 2D peak and $\lambda_M$ in monolayer graphene on hBN. Specifically, the FWHM of the 2D peak remains nearly constant at about 21 cm$^{-1}$ for $\theta>$ 2°, while for $\theta<$ 2° a pronounced broadening occurs, scaling approximately linearly with $\lambda_M$. This broadening is attributed to an inhomogeneous strain distribution with the same periodicity as the moiré potential (see Supplementary Note 3 for further details).
Raman spectroscopy can thus serve as a `fingerprint' for the formation and characteristics of moiré patterns in graphene/hBN, enabling the rapid and non-destructive identification of aligned or misaligned structures.
According to these studies, a broadening of the 2D peak is expected in regions where a moiré superlattice forms due to enhanced interlayer coupling and periodic strain. 
To assess the structural changes induced by this actuation mechanism, we performed spatially resolved Raman spectroscopy on the six samples before and after rotation.

First, focusing on representative sample S1, Figure \ref{fig:S1}d and Figure \ref{fig:S1}e compare the FWHM maps of the 2D Raman peak before and after electrostatic actuation, respectively.
Two distinct regions are clearly visible: a graphene-only area (denoted as `Area A') and a region where graphene is covered by the hBN flake (denoted as `Area B'). In the graphene-only region, the FWHM remains uniform and does not exhibit significant variation after actuation. In contrast, the overlap region displays a homogeneous increase in FWHM following the rotation. This spatial modulation in the overlapping region of the heterostructure suggests a selective change in the vibrational response induced by the twist only where the two layers are in contact.
These observations are corroborated by the representative Raman spectra shown in Figure \ref{fig:S1}g and Figure \ref{fig:S1}h. Each representative spectrum was baseline-corrected; the 2D-band intensity was normalized to the G-band intensity; all spectra belonging to the same region (pre-rotation Area A and Area B, post-rotation Area A and Area B) were then averaged and fitted with a Voigt profile to extract the peak center position ($x_C$) and FWHM (full details in \nameref{sec11}). In Area A, the fit resulted in $x_C$ = 2686.81 $\pm$ 0.23 cm$^{-1}$ and FWHM = 30.22 $\pm$ 0.59 cm$^{-1}$ before rotation. After actuation, $x_C$ was found to be 2686.70 $\pm$ 0.14 cm$^{-1}$, with a corresponding FWHM of 30.52 $\pm$ 0.47 cm$^{-1}$, indicating no appreciable variation relative to the pre-actuation state. In fact, these variations fall within the experimental uncertainty and were not found to be statistically significant (t = 2.45, p = 0.135 for FWHM; t = 2.12, p = 0.170 for $x_C$), indicating that the graphene lattice outside the heterostructure region was unaffected by the induced torque (more details in Supplementary Information).
On the other hand, in the graphene/hBN overlap region, the Raman response changes significantly. Before electrostatic actuation, $x_C$ = 2689.30 $\pm$ 0.24 cm$^{-1}$ with a FWHM of 21.26 $\pm$ 0.66 cm$^{-1}$. After the hBN flake underwent a CCW rotation of approximately 3°, $x_C$ shifts to 2690.90 $\pm$ 0.22 cm$^{-1}$ and broadens to 25.34 $\pm$ 0.63 cm$^{-1}$. In this case, both the shift in $x_C$ and the increase in FWHM are statistically significant (t = 19.24, p = 0.0042 for position; t = 10.06, p = 0.0185 for FWHM), confirming a twist-induced perturbation of the graphene phonon structure due to enhanced interlayer coupling.
The observed FWHM broadening is consistent with the formation of a moiré superlattice, arising from a small-angle rotation between the graphene and hBN lattices.

\noindent Using the empirical relation \cite{eckmann2013raman}:
\begin{equation}
    \text{FWHM}(2D) \simeq 5 + 2.6 \lambda_M
    \label{eq_lambda}
\end{equation}
(with $\lambda_M$ in nm), the post-actuation FWHM of $\sim$25.3 cm$^{-1}$ corresponds to a moiré wavelength of approximately 7.8 nm, consistent with a twist angle $\theta <$ 2° (see Supplementary Information for details). Conversely, the pre-actuation FWHM of $\sim$21.3 cm$^{-1}$ suggests $\theta >$ 2°, thus in line with $\Delta \theta \sim$ 3° inferred from SEM analysis. These estimates support the conclusion that the system transitions into a configuration with enhanced interlayer coupling following the electrostatic rotation.

Similarly, for Sample S2, SEM imaging shows a CW rotation of the hBN flake by $\sim$2°, as shown in Figure \ref{fig:S2}a and Figure \ref{fig:S2}b (see Supplementary Figure S6 for additional characterization of Sample S2). The relative displacement of the flake with respect to the underlying graphene layer is clearly visible in Figure \ref{fig:S2}c and Figure \ref{fig:S2}f, and the presence of a crack near the center provides a reference marker for tracking the angular shift, further confirming the occurrence of interlayer motion following electrostatic actuation. 
The corresponding FWHM maps of the 2D Raman peak, shown in Figure \ref{fig:S2}d and Figure \ref{fig:S2}e, reveal the evolution across the heterostructure before and after rotation. As in Sample S1, two regions can be distinguished: the bare graphene area (Area A) and the graphene/hBN overlap (Area B). Again, in the graphene-only region, the FWHM remains uniform and shows no appreciable change after actuation. In contrast, a clear increase in FWHM is observed in the overlap region. Notably, an inversion in the relative FWHM values occurs: while in the pre-actuation condition the FWHM was larger in the bare graphene compared to the overlap region, after actuation the FWHM becomes higher in the overlap area. This inversion indicates a localized modification of the vibrational properties induced by interlayer rotation.
This behavior is confirmed by the representative Raman spectra extracted from both regions, shown in Figure \ref{fig:S2}g and Figure \ref{fig:S2}h. In Area A, $x_C$ = 2686.94 $\pm$ 0.24 cm$^{-1}$ with a FWHM of 29.20 $\pm$ 0.69 cm$^{-1}$ before actuation, and $x_C$ = 2687.56 $\pm$ 0.16 cm$^{-1}$ with a FWHM of 31.29 $\pm$ 0.47 cm$^{-1}$ after actuation. The variations in both peak position and linewidth are limited, and statistical analysis confirms that these differences are not significant (p = 0.784 for FWHM; p = 0.974 for peak position), indicating that the graphene lattice in the uncovered region remains unaffected by the electrostatic rotation.
In contrast, the graphene/hBN overlap region shows more substantial spectral modifications. The 2D peak shifts from 2689.01 $\pm$ 0.16 cm$^{-1}$ to 2694.41 $\pm$ 0.23 cm$^{-1}$, and the FWHM increases from 24.12 $\pm$ 0.44 cm$^{-1}$ to 32.73 $\pm$ 0.74 cm$^{-1}$ following the $\Delta \theta \sim$ 2° rotation. These changes are statistically significant (t = 4.86, p = 0.040 for peak position; t = 4.48, p = 0.046 for FWHM), confirming a twist-induced modulation of the graphene phonon spectrum due to interlayer reconfiguration.
The observed FWHM broadening and peak shift in the overlap region are consistent with the formation of a quasi-aligned moiré superlattice, which introduces periodic strain within the graphene layer. This interpretation is further supported by the appearance of asymmetry in the 2D peak lineshape after actuation, and by the absence of spectral changes in the uncovered graphene region. Using the empirical relation (\ref{eq_lambda}), the measured FWHM values of 24.12 cm$^{-1}$ and 32.73 cm$^{-1}$ correspond to moiré wavelengths of approximately 7.35 nm and 10.67 nm, respectively. The increase in $\lambda_M$ after actuation reflects a reduction in the twist angle $\theta$, consistent with the transition toward a more aligned stacking configuration. While Raman spectroscopy provides access to the absolute twist angle $\theta$, the SEM analysis yields only the relative angular displacement $\Delta \theta$ between the pre- and post-actuation states.
Since each measured moiré wavelength $\lambda_M$ corresponds to a specific twist angle $\theta$, determining the sign and magnitude of a change $\Delta\theta$ requires adopting a consistent angular convention or relying on additional experimental indicators. Raman spectroscopy yields a twist change of $\Delta\theta \sim 2.42^\circ$, in good agreement with the $\sim 2^\circ$ CW rotation estimated from SEM images using the image-based analysis described previously, considering the combined uncertainties of both methods (see Supplementary Note 3 for further details).

While the experiments demonstrated the feasibility of inducing twist in vdW heterostructures via localized electrostatic torque, we observed important limitations in precision, reversibility, and angular control. Although the system could be actively driven out of its initial state, the induced motion was irreversible. Once a new angular configuration was reached, it remained pinned and could not be driven back to the initial state by further electron beam actuation. Under repeated actuation, we observed no further reconfiguration. In particular, the direction of rotation is consistently governed by the in-plane component of the electric field arising from the applied potential difference, whereas the final angular configuration is set by the energy landscape encountered during rotation, namely by the rotor’s interaction with the moiré potential landscape. We attribute this behavior to a transition from a metastable to a stable state, in which the rotor becomes trapped in the first accessible local energy minimum. In addition, the process was non-deterministic, as the final twist angle could not be predictably controlled by adjusting \( V_\text{bias} \) alone. This behavior may result from the interplay between the applied bias and experimental parameters that coevolve during irradiation, including, for example, the local charge distribution induced by the SEM electron beam, the presence of metal contacts, local variations in hBN thickness and permittivity that modulate the electric field, edge orientation, adhesion and strain, and trapped charges or adsorbates at the interfaces. Since the electrostatic torque evolves dynamically under these coupled effects, the system likely settles into the first energetically accessible configuration rather than a bias-defined target angle.
These observations are consistent with theoretical and experimental studies, which have shown that rotated 2D material heterostructures exhibit complex energy landscapes arising from the interplay of interlayer vdW forces and moiré superlattice formation \cite{zhu2019controlling, zhang2025universal, halbertal2021moire}. These landscapes are characterized by a discrete set of preferred twist angles, often corresponding to commensurate or quasi-commensurate stacking configurations, toward which the system can spontaneously relax. Once such a configuration is reached, the interlayer friction and elastic relaxation combine to stabilize the system in a stable state, which cannot be reconfigured without overcoming substantial energy barriers \cite{halbertal2021moire}.
In our samples, this phenomenon manifests as a spontaneous locking of the hBN layer at a finite rotation angle following charge-induced actuation. Overall, while at the present stage this technique is still subject to the described limitations, at the same time it highlights that charge-injection-induced rotation is already capable of overcoming interlayer friction and stabilizing new angular configurations, laying the groundwork for future developments toward more precise and reversible control.

\section*{Conclusion and Perspectives}
In this work, we have shown that electron-beam-induced charge injection can generate lateral electrostatic fields strong enough to overcome the static friction at the interface between monolayer graphene and a mechanically decoupled hBN flake, resulting in partially controlled in-plane rotation. The effect was observed across six devices, with angular displacements of a few degrees. Three rotated clockwise and three counterclockwise, in line with the bias configuration used during actuation and consistent with an electrostatic origin. SEM images allowed us to track the angular motion while spatially resolved Raman spectroscopy provided an independent signature of interlayer reconfiguration through measurable broadening of the graphene 2D band. Actuation is contactless on the rotor, with the SEM serving as an external trigger for a pre-assembled rotor–stator stack. In addition, the proposed approach does not require integrating micro-electromechanical components. The observed motion is, however, irreversible under current conditions. Once the flake settles into a new angular configuration, it remains pinned, likely due to energy minima in the moiré potential landscape. Looking ahead, we anticipate that this actuation principle can be implemented in a fully on-chip, integrated architecture, by adding a top contact connected to the stator via vertical feedthroughs, enabling the generation of lateral electric fields by applying a small bias directly across the structure. 
Additional improvements may include segmenting the stator into multiple addressable regions to tune the torque distribution and achieve finer angular control. Extending the concept to other materials, such as 2D transition metal dichalcogenides, will broaden the range of twist-sensitive physical responses that could be accessed and dynamically tuned. In summary, this work demonstrates that internal electrostatic forces can be harnessed to induce rotational motion in 2D heterostructures. Integrating this concept into a fully lithographic platform using via-hole contacts may offer a practical route toward reconfigurable devices where the twist degree of freedom becomes a functional parameter, enabling new approaches in optoelectronics, photonics, and quantum materials.
 
\section*{Methods}\label{sec11}
\subsection*{Sample preparation}
CVD graphene was synthesized on commercial copper foil (25 $\mu$m thick, Alfa Aesar, 99.8\% purity, purchased in 2022). To promote the formation of large copper grains, the foils were pre-annealed in a three-zone tube furnace (HZS, Carbolite) in two steps: first at 400 °C for 2–4 h, then at 1060 °C for 2-12 h under an H$_2$/Ar mixture (20 sccm H$_2$, 200 sccm Ar) at $\sim$1 mbar. Following annealing, the foils were electropolished in 70\% H$_3$PO$_4$ at 2.2 V until the current decreased to less than half of its initial value. The copper foils were then cleaned by sequential sonication in acetone (5 min), rinsing with isopropanol (IPA), immersion in HNO$_3$ (30 min), two rounds of DI-water sonication (1 min each), immersion in ethanol (1 min), and drying with N$_2$ gas. The cleaned foils were reloaded into the same quartz tube furnace and the temperature was set to 1000 °C under a H$_2$/Ar mixture (20 sccm H$_2$, 200 sccm Ar) at $\sim$1 mbar. After 75 min, graphene growth was initiated using CH$_4$ (0.04 sccm) for 35 min at 1000 °C and 110 mbar, subsequently cooled by opening the temperature shielding. 
Subsequently, a 50K PMMA resist (Allresist) was spin-coated onto the graphene/copper surface at 3000 rpm for 40 s. Reactive ion etching (Oxford PlasmaPro NPG 80, 30 sccm oxygen and 15 sccm argon, 25 W RF power for 30 s) was employed to remove graphene from the backside of the copper foil. 
The graphene transfer process was carried out as reported in Supplementary S1(b) of \cite{schmuck2022method}. A PMMA-protected graphene sheet on copper substrate is placed on the surface of a Transene copper etchant (CE-100) for 60 min. Once the etching is completed, graphene is transferred with a spoon in sequence to two deionized water baths and a 1:3 solution of HCl 37\%: deionized water to remove traces of copper etchant, and finally to a third deionized water bath. From this last water bath, the graphene sheet is transferred on the Si/SiO$_2$ substrate by fishing, and let dry inside an upside down beaker to avoid surface contamination, with an inclination of 30° - 45°, for 1.5 hours. Afterwards, the Si/SiO$_2$/graphene sample is annealed first at 80°C for 30 min in air, and then at 80°C in vacuum over-night to remove water residues and increase the adhesion of graphene. The protective PMMA layer on top of graphene is then removed by first placing the samples in acetone for 10 minutes, and then boiling acetone at 54°C for 1 h. The sample is finally rinsed in isopropyl alcohol and blown dried with a nitrogen gun.

Hexagonal boron nitride flakes with a nominal thickness of approximately 30 nm were obtained via mechanical exfoliation from bulk hBN crystals (HQ Graphene). The exfoliation process was performed using a standard adhesive tape-based method (Scotch Magic Tape 810, 3M). A fresh surface of the bulk crystal was first brought into contact with the adhesive side of a tape and subjected to repeated folding and unfolding cycles to progressively thin down the material. The tape carrying the exfoliated flakes was then gently pressed onto a clean SiO$_2$/Si substrate (with a 285 nm thick thermally grown SiO$_2$ layer) under ambient conditions. After ensuring conformal contact, the tape was slowly peeled off, leaving behind a distribution of hBN flakes on the substrate surface. Optical microscopy was used to locate candidate flakes, and the thickness was estimated based on optical contrast and subsequently confirmed by atomic force microscopy in tapping mode. All substrates were cleaned in acetone and isopropanol and dried under nitrogen flow prior to exfoliation to minimize contamination and improve flake adhesion. A transparent elastomer stamp was prepared by drop-casting poly(propylene carbonate) (PPC, Sigma-Aldrich, CAS 25511-85-7) onto a PDMS layer supported by a clean glass slide. The assembly was briefly heated to $\sim$60°C to ensure uniform spreading and adhesion of the PPC to the PDMS surface. Using a micromanipulator-based transfer stage (HQ Graphene), the PPC/PDMS stamp was aligned and brought into contact with the selected hBN flake at a substrate temperature of $\sim$40°C. The flake adhered to the PPC layer and was lifted from the substrate. The hBN flake was aligned over the target graphene and brought into contact using the same transfer setup. Heating the stage to $\sim$90°C promoted adhesion between hBN and graphene, after which the stamp was retracted, leaving the hBN in place. The remaining PPC residue was removed by immersion in chloroform followed by isopropanol, and the sample was dried under nitrogen flow.

CVD graphene was patterned by first spin-coating a 60-nm-thick layer of PMMA 50k (AR-P 632.06, Allresist GmbH) on it and baking the sample for 5 min at 180°C on a hot plate followed by spin-coating of a 60-nm-thick layer of PMMA 950 k (AR-P 672.02, Allresist GmbH) and a 5 min bake at 180°C on a hot plate. Electron beam lithography was carried out with a Vistec EBPG 5200+ system and the samples were developed by rinsing them in 1:4 MIBK:IPA solution for 60 s followed by a brief rinse in pure IPA and drying with nitrogen gun. The exposed graphene was etched with an Oxford PlasmaPro NPG 80 system using a mixture of oxygen (30 sccm) and argon (15 sccm), RF power of 25 W and an etching time of 8 s. The PMMA was removed by soaking overnight in N-Methyl-2-pyrrolidone (NMP). For patterning the metal electrodes, a 320-nm-thick layer of PMMA/MA 33\% (AR-P 617.06, Allresist GmbH) was spin-coated and baked for 5 min at 180°C on a hot plate followed by spin-coating of an 87-nm-thick layer of PMMA 950 k (AR-P 672.02, Allresist GmbH) and a 5 min bake at 180°C on a hot plate. Electron beam lithography patterning and developing were then carried out as in the case of graphene etching and 5 nm titanium and 60 nm gold were evaporated on the sample using an Evatec BAK501 LL electron beam evaporator. Finally, a lift-off of the metal was carried out in DMSO heated to 80°C, followed by a rinse in IPA and drying of the samples with a nitrogen gun.

\subsection*{Atomic Force Microscopy}
AFM imaging and thickness metrology were performed on a Park Systems XE-100 (Park Systems, Korea) in intermittent-contact (tapping) mode using PPP-NCHR silicon cantilevers (NanoWorld AG, Neuchâtel, Switzerland; nominal tip radius $<$10 nm; resonance $\sim$330 kHz). Candidate hBN flakes were pre-selected by optical contrast on 285 nm SiO$_2$/Si substrates. Images (512 $\times$ 512 pixels) were acquired at a scan rate of 0.2 Hz, keeping the setpoint $>$ 65\% of the free oscillation amplitude. Height profiles were extracted across flake edges to determine step height relative to the substrate; the samples were analyzed with Gwyddion 2.53. Exfoliated hBN thicknesses typically ranged 28–32 nm.

\subsection*{Scanning Electron Microscopy}
SEM imaging was performed on a JEOL JSM-6490LA (JSM-6500 series; JEOL Ltd., Japan) using a secondary-electron detector with an accelerating voltage of 10 kV and a beam current of 80 pA.
For actuation, the SEM was operated at 5 kV to minimize beam damage. Beam parameters were calibrated to deliver a sufficient electron dose to induce rotation while preserving crystal quality. Full details and dose estimates are reported in the Discussion and Supplementary Information. In-situ electrical probing was carried out inside the SEM using an Imina Technologies micromanipulation platform (Plan-les-Ouates, Switzerland) equipped with tungsten microprobes. A Keithley 2612 SourceMeter (Keithley Instruments, Cleveland, OH, USA) provided the ground reference and, when required, an applied bias to the contacted pads.

\subsection*{Raman Spectroscopy}
Measurements were performed on a Renishaw inVia micro-Raman system (Renishaw plc, Wotton-under-Edge, UK; 532 nm excitation) focused with a 100$\times$, NA = 0.9 objective. The spectrometer used a 2400 lines/mm grating. Incident power at the sample was 1 mW and the integration time was 1 s per spectrum. Raman mapping was performed on the graphene and graphene/hBN overlap areas of the samples before and after SEM treatment using identical settings. For each spectrum, the 2D band was fitted with a Voigt profile to extract peak position and FWHM.

\subsection*{Twist Angle Extraction from SEM Images}
To quantitatively determine the in-plane twist angle induced by electron-beam exposure, we developed a Python routine that performs automated registration between pre- and post-rotation SEM images.
The images are converted to 8-bit grayscale and cropped around a fixed region of interest (ROI), typically around 80\% of the total field of view, centered on the lithographically defined metal cross deposited on the hBN layer.
The algorithm performs a two-stage rigid rotation search: a coarse scan over the interval [-45°, 45°] with 0.05° steps to identify a preliminary minimum, followed by a fine scan within a $\pm$0.3° window using 0.002° steps.
For each trial angle, the post-rotation image is rotated and the mean squared error (MSE) with respect to the pre-rotation image is calculated over the ROI. The twist angle is defined as the angle that minimizes the MSE, while the uncertainty is estimated as the half-width at which the MSE increases by 1\%. The code is provided in the Supplementary Data.

\backmatter



\bmhead{Funding}

N.C. discloses support for the research of this work from the European Union’s Horizon Europe research and innovation programme under the Marie Skłodowska-Curie grant agreement No. 101109662.

\bmhead{Acknowledgements}

The authors thank the Electron Microscopy Facility and the Material Characterization Facility of the Italian Institute of Technology for support with material characterization, as well as the support provided by the Clean Room Facility of the Italian Institute of Technology and the staff of the Binnig and Rohrer Nanotechnology Center (BRNC) for device fabrication.





\nolinenumbers


\newpage

{\huge\textbf{Supplementary Information}\par}


\section{Device concept}
\begin{figure}[ht]
    \centering
    \includegraphics[width=\linewidth]{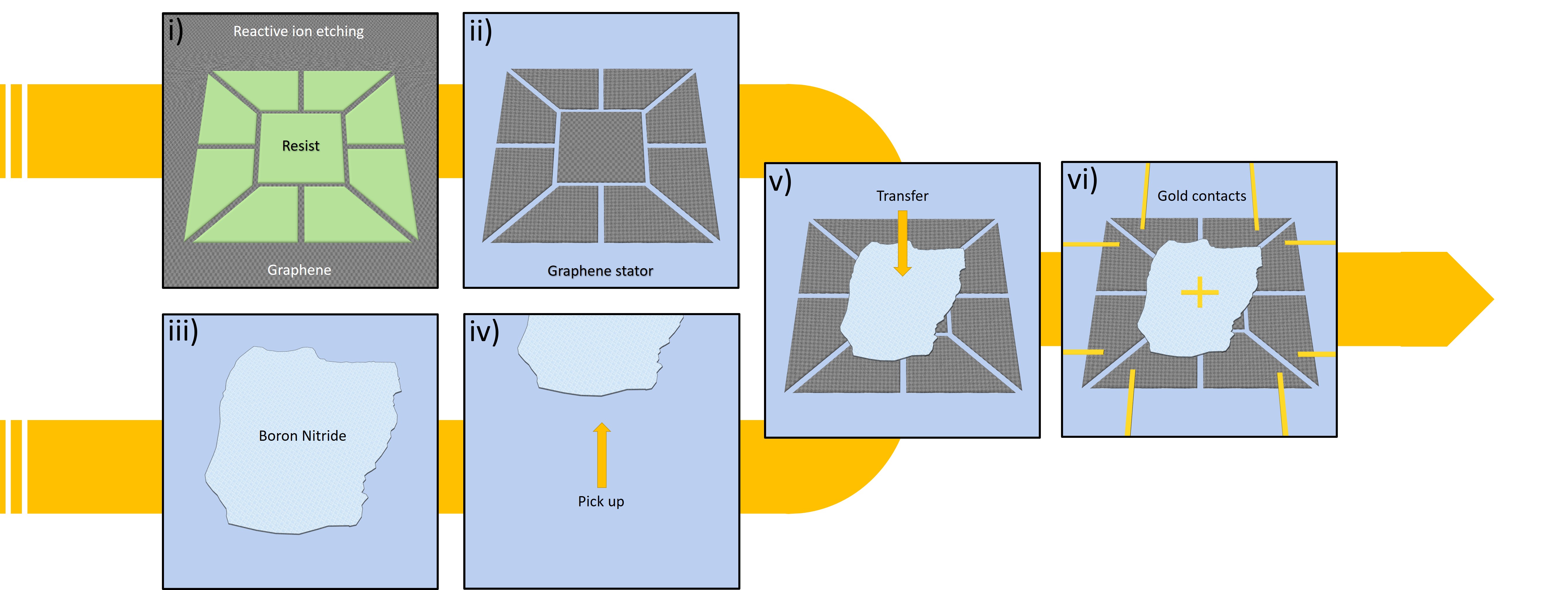}
    \caption{Step-by-step fabrication process of the twistable graphene/hBN heterostructure.  
    i) Patterning of monolayer graphene via electron beam lithography followed by reactive ion etching (RIE), using a resist mask.  
    ii) Final geometry of the graphene stator after RIE.  
    iii) Exfoliation of a hexagonal boron nitride (hBN) flake onto a separate substrate.  
    iv) Pick-up of the hBN flake using a transparent polymer stamp.  
    v) Dry-transfer of the hBN flake onto the patterned graphene stator.  
    vi) Definition of gold contacts via e-beam lithography and thermal evaporation.}
    \label{fig:figS0}
\end{figure}
The full fabrication sequence of the graphene/hBN heterostructure used for twist control experiments is illustrated in Supplementary Figure \ref{fig:figS0}. The process begins with the growth of high-quality monolayer graphene via low-pressure chemical vapor deposition (CVD) on copper foil, followed by a standard PMMA-assisted transfer onto a Si/SiO$_2$ substrate.

To define the stator geometry, a resist layer (i.e., PMMA) is first spin-coated onto the graphene surface, and electron beam lithography (EBL) is used to pattern a central square region surrounded by eight triangular segments extending toward the edges of the chip (panel i). The exposed graphene is then etched via reactive ion etching (RIE) under O$_2$/Ar plasma conditions to remove unprotected areas, yielding the final etched geometry shown in panel ii. The geometry was specifically designed to enable control over the direction of the lateral electric field generated during SEM irradiation. By varying the position of grounded pads relative to the electron beam, it is possible to steer the in-plane torque vector acting on the hBN flake. At the same time, this layout maximizes the suspended graphene area, ensuring that the hBN rests entirely on the stator without substrate interaction. In fact, the central square region is completely isolated electrically and serves a purely mechanical function and acts as a suspended support for the hBN flake. Its role is to prevent the hBN from coming into contact with the underlying SiO$_2$ substrate, thereby avoiding flake locking due to friction or interfacial strain. The parallelepiped-shaped pads, in contrast, are later connected to gold electrodes and used for electrostatic grounding. Separately, a multilayer hBN flake ($\sim$30 nm thick) is exfoliated from a bulk crystal (HQ Graphene) using the standard Scotch tape method and deposited onto a sacrificial SiO$_2$/Si chip (panel iii). Candidate flakes are identified by optical contrast and later verified by atomic force microscopy to confirm thickness uniformity and surface quality. The selected hBN flake is then picked up using a transparent elastomeric stamp consisting of polypropylene carbonate (PPC) on PDMS (panel iv), mounted on a micromanipulator-controlled dry-transfer stage. The stamp is aligned with the central stator region and brought into gentle contact at $\sim$90 °C (panel v), enabling adhesion of the flake to the graphene stator. The stamp is then retracted, and residual PPC is removed in chloroform and rinsed in isopropanol. Finally, gold electrodes are defined by a second EBL step followed by thermal evaporation and lift-off (panel vi). These electrodes serve to ground selected graphene regions during SEM irradiation, allowing for well-controlled and asymmetric electrostatic boundary conditions.

\section{Energy dose extimation}

Below is a step-by-step estimation of the energy-dose provided by SEM.

\subsubsection*{Experimental Parameters}
\begin{itemize}
  \item \textbf{Accelerating voltage}, \(V\): \(5\,\mathrm{keV}\)
  \item \textbf{Beam current}, \(I\): up to \(100\,\mathrm{pA} = 1.0\times10^{-10}\,\mathrm{A}\)
  \item \textbf{Exposure (scan) time}, \(t\): \(1\,\mathrm{s}\)
  \item \textbf{Irradiated area}, \(A\): \(50 \times 50\,\mu\mathrm{m}^2 = 2.5 \times 10^{-9}\,\mathrm{m}^2\)
\end{itemize}

\subsubsection*{Energy per Electron}
Converting \(5\,\mathrm{keV}\) to joules:
\begin{equation}
  E_e \;=\; 5\,\mathrm{keV}
         \;=\; 5 \times 1.602 \times 10^{-16}\,\mathrm{J}
         \;\approx\; 8.01 \times 10^{-16}\,\mathrm{J/e}^-.
\end{equation}

\subsubsection*{Total Energy Delivered in 1\,s}
\paragraph{(a) Charge delivered, \(Q\):}
\begin{equation}
  Q \;=\; I \times t
        \;=\; (1.0 \times 10^{-10}\,\mathrm{A}) \times (1\,\mathrm{s})
        \;=\; 1.0 \times 10^{-10}\,\mathrm{C}.
\end{equation}

\paragraph{(b) Number of electrons, \(N\):}
\begin{equation}
  N \;=\; \frac{Q}{e}
         \;=\; \frac{1.0 \times 10^{-10}\,\mathrm{C}}{1.602 \times 10^{-19}\,\mathrm{C/e}^-}
         \;\approx\; 6.24 \times 10^8 \,\mathrm{electrons}.
\end{equation}

\paragraph{(c) Total energy, \(E_{\text{tot}}\):}
\begin{equation}
  E_{\text{tot}} \;=\; N \times E_e
                   \;=\; (6.24 \times 10^8) \times (8.01 \times 10^{-16}\,\mathrm{J})
                   \;\approx\; 5.0 \times 10^{-7}\,\mathrm{J}.
\end{equation}

\subsubsection*{Energy Density}
Given the exposed area \( A = 2.5 \times 10^{-9}\,\mathrm{m}^2\):
\begin{equation}
  E_{\mathrm{dens}}
    \;=\; \frac{E_{\mathrm{tot}}}{A}
    \;=\; \frac{5.0 \times 10^{-7}\,\mathrm{J}}{2.5 \times 10^{-9}\,\mathrm{m}^2}
    \;\approx\; 2.0 \times 10^2\,\mathrm{J\,m}^{-2}.
\end{equation}
This corresponds to \(\sim 200\,\mathrm{J\,m}^{-2}\) per 1\,s scan at \(100\,\mathrm{pA}\). Even if we account for variations in beam current, multiple scans, or partial overlap during scanning, the resulting dose remains in the \(10^2\)--\(10^3\,\mathrm{J\,m}^{-2}\) range.

\section{Raman-based estimation of moiré periodicity and twist angle in graphene/hBN heterostructures}

In monolayer graphene, the Raman spectrum typically exhibits three prominent peaks: the G peak, the 2D peak, and the D peak. The 2D peak, centered around 2700 cm$^{-1}$, arises from a second-order process involving two phonons near the K point of the Brillouin zone and is highly sensitive to the number of layers and stacking order \cite{ferrari2013raman}.
In graphene/hBN heterostructures, the superposition of two periodic lattices with a slight lattice mismatch or relative rotation generates a long-wavelength moiré pattern. This periodic potential can modulate both the electronic structure and the local strain within the graphene layer. These strain variations, periodic in nature, are reflected in the Raman response through broadening and asymmetry of the 2D peak, associated with inhomogeneous phonon confinement and local variations in interlayer coupling.
Several studies have shown a correlation between Raman spectral features and the moiré wavelength ($\lambda_M$) in graphene/hBN heterostructures \cite{eckmann2013raman,cheng2015raman, finney2019tunable}. Notably, Eckmann et al. \cite{eckmann2013raman} reported a linear relationship between the full width at half maximum (FWHM) of the 2D peak and $\lambda_M$ in monolayer graphene on hBN. In particular, the variation in the FWHM of the 2D band in Raman spectroscopy for misalignment angles ($\theta$) between graphene and hBN behaves differently depending on whether $\theta$ is greater or less than 2°. For $\theta>$ 2° (misaligned structures), the FWHM(2D) is about 21 cm$^{-1}$. Instead, for $\theta<$ 2° (slightly misaligned structures or near alignment), a characteristic and pronounced broadening of the FWHM(2D) is observed as $\theta$ decreases. 
This broadening is attributed to an inhomogeneous strain distribution with the same periodicity as the moiré potential. 
This phenomenology can be quantitatively described by the geometric relation between $\lambda_M$ and $\theta$, and by the empirical correlation between $\lambda_M$ and the FWHM(2D). 

According to Yankowitz et al., \cite{Yankowitz2012SuperlatticeDirac} the relationship between the moiré superlattice periodicity, $\lambda_M$, and the twist (or misalignment) angle, $\theta$, in graphene/hBN systems is given by:

\begin{equation}
\lambda_M = \frac{a_{CC}}{\sqrt{2(1+\delta)(1-\cos\theta) + \delta^2}},
\end{equation}

where
\[
a_{CC} = \SI{0.246}{nm} \quad \text{(the graphene lattice parameter)}
\]
and
\[
\delta \approx 0.018 \quad \text{(the lattice mismatch between graphene and hBN)}.
\]

\subsection*{Empirical Relation with Raman 2D Peak}

The FWHM of the 2D Raman peak is empirically related to the moiré wavelength by \cite{eckmann2013raman}:
\[
\text{FWHM(2D)} \simeq 5 + 2.6\, \lambda_M,
\]
with FWHM in \si{\per\centi\meter} and $\lambda_M$ in \si{nm}.

\subsection*{Calculation}

\paragraph{Case 1: FWHM(2D) = \SI{24.12}{cm^{-1}}}

\begin{enumerate}
    \item \textbf{Calculate $\lambda_M$:}
    \[
    \lambda_M = \frac{24.12 - 5}{2.6} \approx \SI{7.35}{nm}.
    \]
    
    \item \textbf{Invert the Moiré periodicity equation to solve for $\theta$:} \\
    Define
    \[
    X = \frac{a_{CC}}{\lambda_M} \approx \frac{0.246}{7.35} \approx 0.0335,
    \]
    so that
    \[
    X^2 \approx 0.00112, \quad \delta^2 \approx 0.000324.
    \]
    
    Rearranging the periodicity equation:
    \[
    1 - \cos\theta = \frac{X^2 - \delta^2}{2(1+\delta)} \approx \frac{0.00112 - 0.000324}{2 \times 1.018} \approx 0.000391.
    \]
    
    Then,
    \[
    \theta \approx \arccos(0.999609) \approx \SI{0.028}{rad} \approx \SI{1.6}{^\circ}.
    \]
\end{enumerate}

\paragraph{Case 2: FWHM(2D) = \SI{32.73}{cm^{-1}}}

\begin{enumerate}
    \item \textbf{Calculate $\lambda_M$:}
    \[
    \lambda_M = \frac{32.73 - 5}{2.6} \approx \SI{10.67}{nm}.
    \]
    
    \item \textbf{Determine $\theta$:} \\
    Compute
    \[
    X = \frac{0.246}{10.67} \approx 0.02304, \quad X^2 \approx 0.000532.
    \]
    
    Then,
    \[
    1 - \cos\theta \approx \frac{0.000532 - 0.000324}{2 \times 1.018} \approx 0.000102,
    \]
    leading to
    \[
    \theta \approx \arccos(0.999898) \approx \SI{0.0143}{rad} \approx \SI{0.82}{^\circ}.
    \]
\end{enumerate}

\paragraph{Case 3: FWHM(2D) = \SI{21.26}{cm^{-1}}}

\begin{enumerate}
    \item \textbf{Calculate $\lambda_M$:}
    \[
    \lambda_M = \frac{21.26 - 5}{2.6} \approx \SI{6.25}{nm}.
    \]
    
    \item \textbf{Determine $\theta$:}\\
    Define
    \[
    X = \frac{a_{CC}}{\lambda_M} \approx \frac{0.246}{6.25} \approx 0.03936,
    \]
    hence
    \[
    X^2 \approx 0.00155, \quad \delta^2 \approx 0.000324.
    \]
    
    Then,
    \[
    1 - \cos\theta = \frac{0.00155 - 0.000324}{2 \times 1.018} \approx \frac{0.001226}{2.036} \approx 0.000602.
    \]
    
    Thus,
    \[
    \theta \approx \arccos(0.999398) \approx \SI{0.03468}{rad} \approx \SI{2.0}{^\circ}.
    \]
    
    \textbf{Note:} In the misaligned regime ($\theta > \SI{2}{^\circ}$) the FWHM(2D) tends to saturate around \SI{21}{cm^{-1}}, so deducing the exact twist angle from the Raman broadening is less reliable in this case.
\end{enumerate}

\paragraph{Case 4: FWHM(2D) = \SI{25.34}{cm^{-1}}}

\begin{enumerate}
    \item \textbf{Calculate $\lambda_M$:}
    \[
    \lambda_M = \frac{25.34 - 5}{2.6} \approx \SI{7.82}{nm}.
    \]
    
    \item \textbf{Determine $\theta$:}\\
    Compute
    \[
    X = \frac{0.246}{7.82} \approx 0.03146, \quad X^2 \approx 0.000990.
    \]
    
    Then,
    \[
    1 - \cos\theta \approx \frac{0.000990 - 0.000324}{2 \times 1.018} \approx \frac{0.000666}{2.036} \approx 0.000327.
    \]
    
    Thus,
    \[
    \theta \approx \arccos(0.999673) \approx \SI{0.02557}{rad} \approx \SI{1.46}{^\circ}.
    \]
    
\end{enumerate}

\begin{itemize}
    \item Lower FWHM values (around \SI{21}{cm^{-1}}) typically correspond to larger misalignment angles ($\theta > \SI{2}{^\circ}$). In such cases, as seen for FWHM(2D) = \SI{21.26}{cm^{-1}}, the Raman broadening tends to saturate, and the empirical relation becomes less sensitive, making it difficult to precisely deduce the twist angle.
    \item Higher FWHM values (e.g., \SI{24.12}{cm^{-1}} or \SI{32.73}{cm^{-1}}) are associated with nearly aligned superlattices ($\theta < \SI{2}{^\circ}$), where the empirical relation yields more reliable estimates of the twist angle.
\end{itemize}

The Raman FWHM(2D) method allows estimation of the absolute value of the twist angle between graphene and hBN layers via the derived moiré wavelength. However, it does not inherently provide the twist direction (clockwise versus counterclockwise). By combining measurements with additional information (or by assigning a convention), one can deduce the net relative misorientation. For S2 (i.e., \SI{24.12}{cm^{-1}} and \SI{32.73}{cm^{-1}}) may correspond to twist angles of approximately \SI{1.6}{^\circ} and \SI{0.82}{^\circ}, which, if taken with opposite signs, yield a net twist of about \SI{2.42}{^\circ}.

To assign a sense (CW or CCW), one must adopt a convention or use additional experimental data. For example, one possible assignment is to take +1.6° and –0.82°, obtaining a net twist of:

    \[1.6^\circ-(-0.82^\circ)=2.42^\circ~\text{(CCW)}.
     \]
Alternatively, if both angles are taken with the same sign (for example, both as positive), then the difference is:
   \[1.6^\circ-0.82^\circ=0.78^\circ~\text{(CW)}.
     \]
     According to the SEM image analysis, the net twist angle is determined to be \(\theta = 2.42^\circ\) CCW.

\section{Extended characterization}

\subsection*{CVD Graphene before patterning}
\begin{figure}[H]
  \centering
  \includegraphics[width=0.65\linewidth]{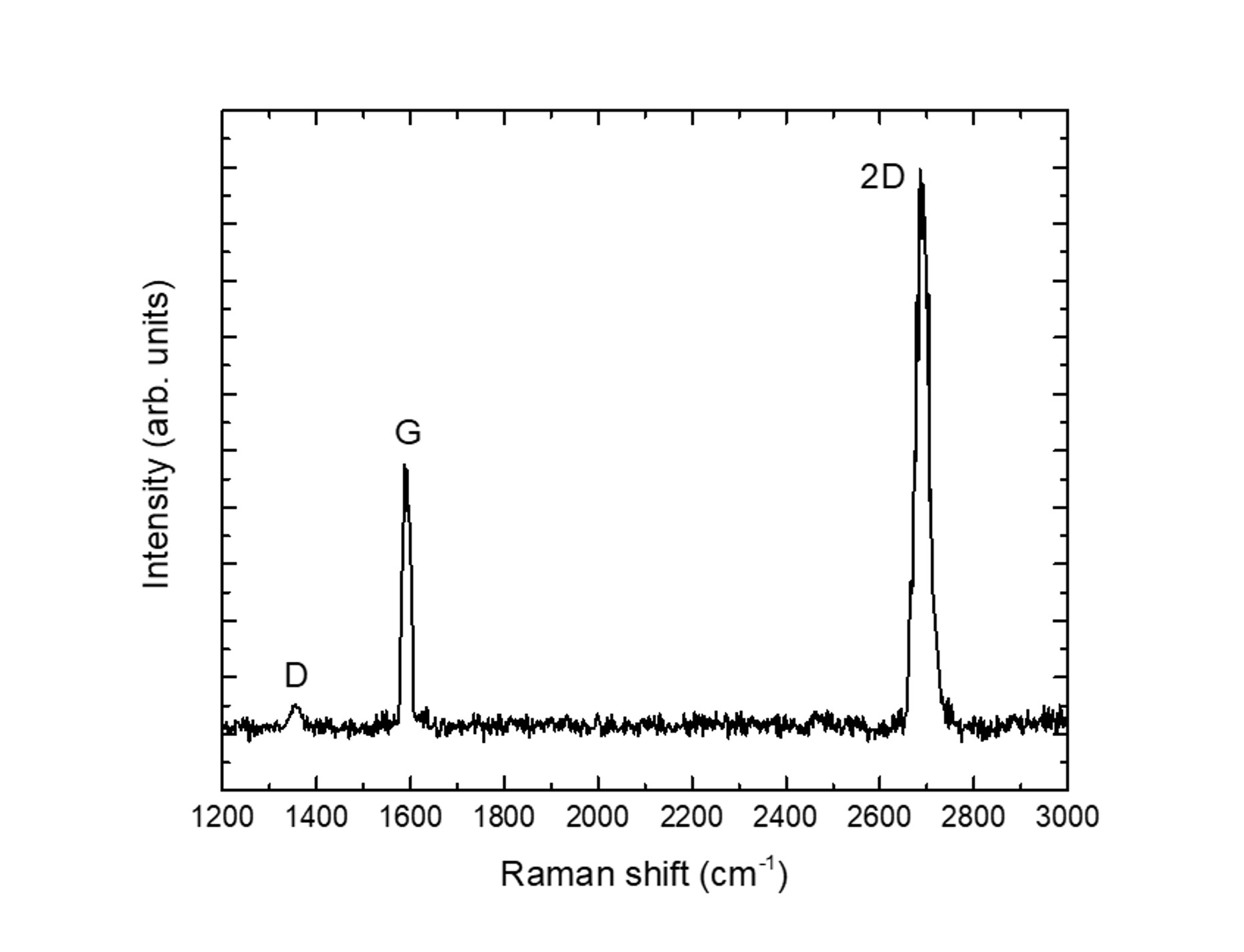}
  \caption{Representative Raman spectrum from the mapped area.}
\end{figure}

Raman spectra were acquired on CVD monolayer graphene transferred onto SiO$_2$/Si. A selected area was mapped, and the spectra were analyzed by fitting the D, G, and 2D bands with Voigt profiles. For each parameter, we report the mean value and the standard deviation calculated over the mapped area. The extracted quantities are the peak center position ($x_C$), the full width at half maximum (FWHM), and the integrated intensity ($I$).

\renewcommand{\arraystretch}{1.4} 
\begin{table}[h]
\centering
\begin{tabular*}{0.7\textwidth}{@{\extracolsep{\fill}}lcc}
\hline
 & Value $\pm$ Std.~Dev. & Units \\
\hline
$x_C(D)$      & $1354.97 \pm 0.80$ & cm$^{-1}$ \\
$x_C(G)$      & $1588.69 \pm 5.83$ & cm$^{-1}$ \\
$x_C(2D)$     & $2689.04 \pm 2.19$ & cm$^{-1}$ \\
\hline
FWHM(D)       & $24.90 \pm 1.42$   & cm$^{-1}$ \\
FWHM(G)       & $16.91 \pm 3.57$   & cm$^{-1}$ \\
FWHM(2D)      & $31.68 \pm 0.83$   & cm$^{-1}$ \\
\hline
$I_D/I_G$     & $0.113 \pm 0.030$  & -- \\
$I_{2D}/I_G$  & $2.117 \pm 0.151$  & -- \\
\hline
\end{tabular*}
\vspace{0.5em}
\end{table}
\renewcommand{\arraystretch}{1.0} 

\noindent The ratio $I_{2D}/I_G > 2$ together with a single 2D band of $\mathrm{FWHM}(2D) \simeq 31.7$~cm$^{-1}$ confirms the monolayer character of CVD graphene. The narrow dispersion of $\mathrm{FWHM}(2D)$ ($\sigma \approx 0.83$~cm$^{-1}$) indicates a limited spatial inhomogeneity of the 2D mode within the mapped area. The consistently low $I_D/I_G$ ratio ($\sim 0.11$) is indicative of a low density of defects. No D$'$ band was resolved above noise under the employed acquisition conditions.

\subsubsection*{Defect density estimation}

From the ratio $I_D/I_G$ we can estimate the inter-defect distance $L_D$ using the Cançado relation in the so-called ``stage-1'' regime~\cite{canccado2011quantifying}:

\begin{equation}
L_D^2 = (1.8 \times 10^{-9}) \, \lambda^4 \, \frac{I_G}{I_D},
\end{equation}

\noindent with $\lambda = 532$~nm. Inserting the fitted parameters gives:

\[
L_D \approx 36 \pm 5~\text{nm}.
\]

\noindent The corresponding defect density is then \cite{canccado2011quantifying}:

\[
n_D = \frac{1}{\pi L_D^2} \approx (2.5 \pm 0.7) \times 10^{10}~\text{cm}^{-2}.
\]

\noindent These values are consistent with a low density of defects.

\subsection*{CVD Graphene after patterning and after electron beam irradiation}
\begin{figure}[H]
  \centering
  \includegraphics[width=0.65\linewidth]{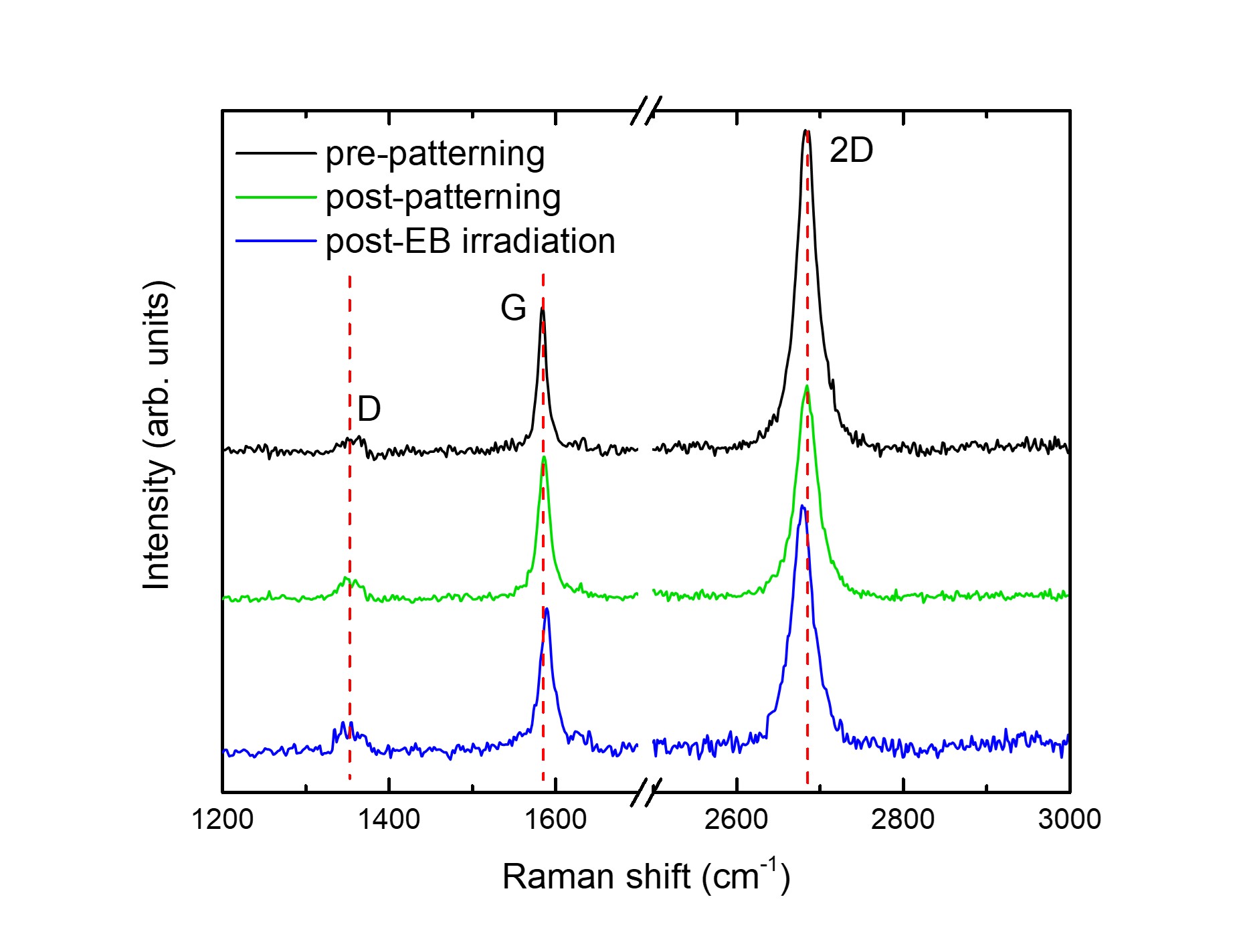}
  \caption{Representative Raman spectra of CVD monolayer graphene acquired before patterning (black), after patterning (green), and after electron-beam (EB) irradiation (blue). Spectra are baseline-corrected and vertically offset for clarity. Red dashed lines indicate the shift in peak positions (\(\mathrm{D}\), \(\mathrm{G}\), \(2\mathrm{D}\)).}
\end{figure}

\begin{figure}[H]
  \centering
  \includegraphics[width=\linewidth]{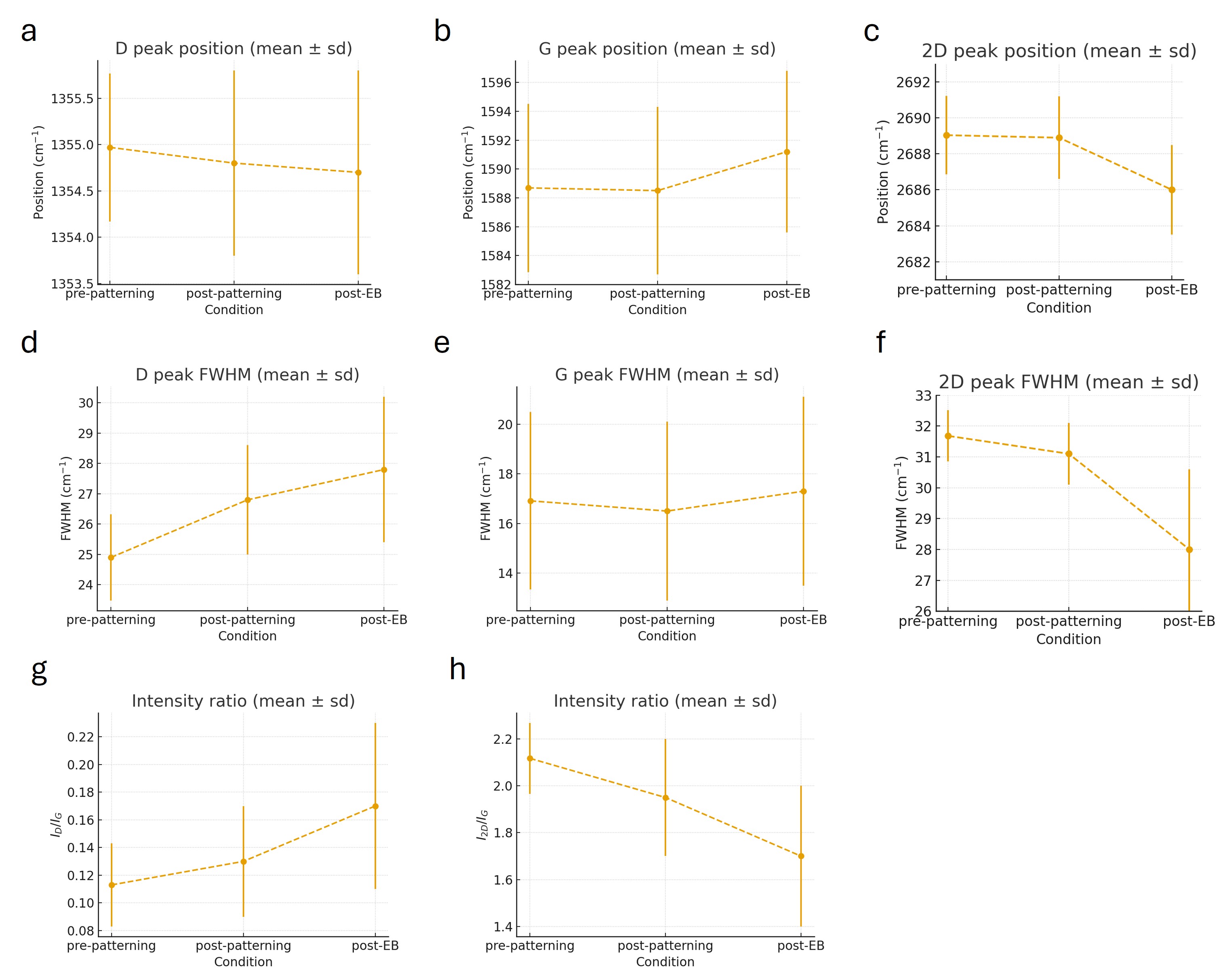}
  \caption{Evolution of Raman parameters for CVD monolayer graphene under three conditions: pre-patterning, post-patterning, and post-EB irradiation. Panels (a–c) show peak center positions \(x_C\) for D, G, and 2D; panels (d–f) show the corresponding \(\mathrm{FWHM}\); panels (g–h) report the intensity ratios \(I_D/I_G\) and \(I_{2D}/I_G\) computed from Voigt-fit. Dashed lines are guides to the eye.}

  \label{panelSI}
\end{figure}

Raman spectra were acquired on different graphene samples before patterning, after patterning, and after electron-beam irradiation. For each condition, statistics (\(\text{mean} \pm \text{sd}\)) were computed and the three conditions were compared (Figure \ref{panelSI}). All spectra were baseline-corrected before the fitting. The center position \(x_C\), the full width at half maximum \(\mathrm{FWHM}\), and the ratios \(I_D/I_G\) and \(I_{2D}/I_G\) were obtained from Voigt fits. Error bars represent one standard deviation across the mapped area.
After patterning, the \(\mathrm{G}\) and \(2\mathrm{D}\) bands show no systematic blue-/red-shift:
\(x_C(\mathrm{G}):\, 1588.69\pm5.83 \to 1588.5\pm5.8~\text{cm}^{-1}\) \((\mathrm{FWHM}:\, 16.91\pm3.57 \to 16.5\pm3.6~\text{cm}^{-1})\);
\(x_C(2\mathrm{D}):\, 2689.04\pm2.19 \to 2688.9\pm2.3~\text{cm}^{-1}\) \((\mathrm{FWHM}:\, 31.68\pm0.83 \to 31.1\pm1.0~\text{cm}^{-1})\).
Intensity ratios: \(I_D/I_G:\, 0.113\pm0.030 \to 0.130\pm0.040\); \(I_{2D}/I_G:\, 2.117\pm0.151 \to 1.95\pm0.25\).

After electron beam irradiation, the G band undergoes a blue-shift with moderate broadening ($x_C$(G): 1588.5 $\pm$ 5.8 $\rightarrow$ 1591.2 $\pm$ 5.6 cm$^{-1}$; FWHM(G): 16.5 $\pm$ 3.6 $\rightarrow$ 17.3 $\pm$ 3.8 cm$^{-1}$), whereas the 2D band shows a red-shift with a net narrowing of the mean linewidth ($x_C$(2D): 2688.9 $\pm$ 2.3 $\rightarrow$ 2686.0 $\pm$ 2.5 cm$^{-1}$; FWHM(2D): 31.1 $\pm$ 1.0 $\rightarrow$ 28.0 $\pm$ 2.6 cm$^{-1}$). This opposite blue-/red-shift correlation, together with the limited increase of $I_{D}/I_G$ (0.130 $\pm$ 0.040 $\rightarrow$ 0.170 $\pm$ 0.060) and the absence of D', indicates charge-doping and reduction in nanometre-scale strain, rather than beam-induced lattice damage \cite{neumann2015raman}.

\subsection*{S1}

\begin{figure}[H]
    \centering
    \includegraphics[width=\linewidth, trim=0 5 10 0, clip]{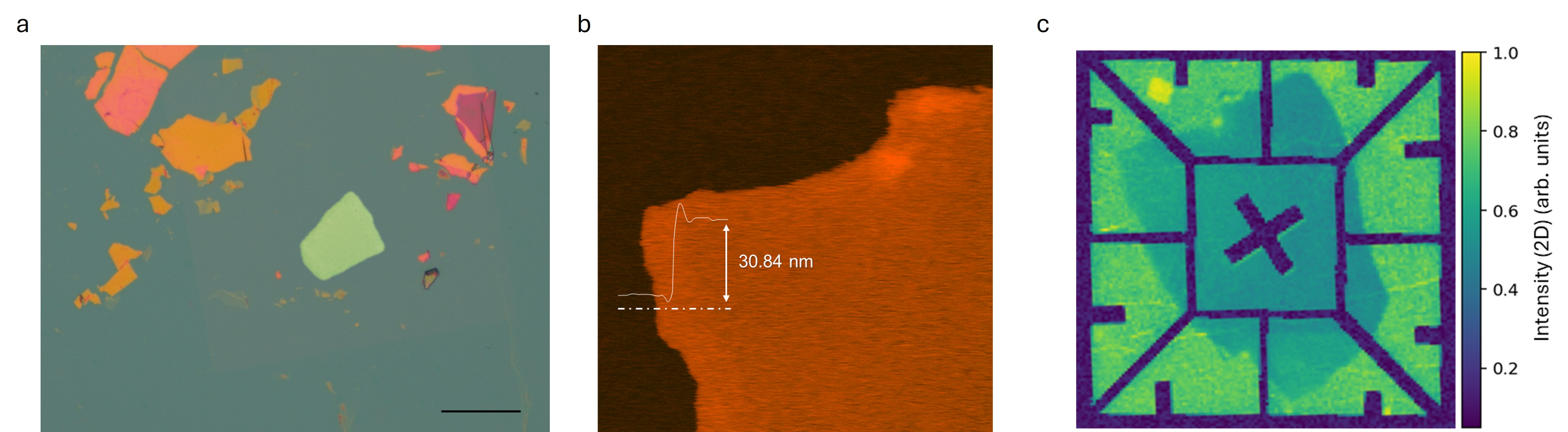}
\caption{
a) Optical micrograph of the exfoliated hBN flake on 285 nm SiO$_2$/Si before the transfer.
b) AFM topography at the flake edge (tapping mode); step height 30.84 nm.
c) Raman map of the graphene 2D-band intensity on the graphene/hBN heterostructure, normalized to the local maximum; regions A (bare graphene) and B (graphene under hBN) are distinguishable by the reduced 2D intensity in the overlap.}
 \label{fig:figS1}
\end{figure}

Sample S1 was characterized by optical microscopy, AFM, and Raman spectroscopy. The hBN flake used to cover graphene was first identified on 285 nm SiO$_2$/Si (Figure \ref{fig:figS1}a). AFM at the flake edge yields a thickness of 30.84 nm (Figure \ref{fig:figS1}b). A Raman map of the graphene 2D-band intensity acquired on the graphene/hBN stack with 532 nm excitation delineates covered and uncovered regions, showing a systematic reduction of the 2D intensity under hBN consistent with optical interference in the multilayer substrate (Figure \ref{fig:figS1}c). 

\subsection*{S2}

\begin{figure}[H]
    \centering
    \includegraphics[width=0.7\linewidth, trim=0 5 10 0, clip]{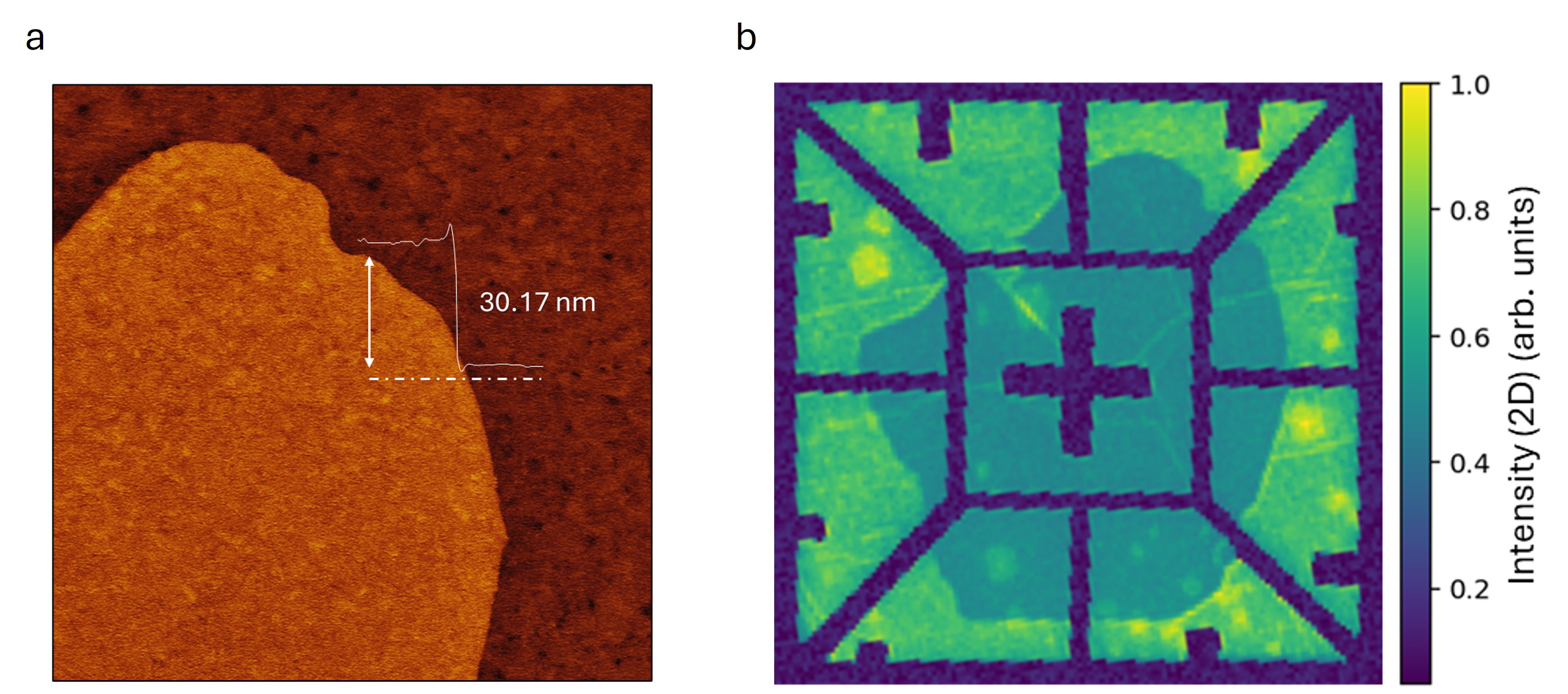}
\caption{
a) AFM topography at the flake edge (tapping mode); step height 30.17 nm.
b) Raman map of the graphene 2D-band intensity on the graphene/hBN heterostructure, normalized to the local maximum; regions A (bare graphene) and B (graphene under hBN) are distinguishable by the reduced 2D intensity in the overlap.}
 \label{fig:figS2}
\end{figure}

Sample S2 was characterized by AFM and Raman spectroscopy. AFM at the hBN edge yields a thickness of 30.17 nm (Figure \ref{fig:figS2}a). A Raman 2D-intensity map acquired on the graphene/hBN heterostructure with 532 nm excitation distinguishes covered from bare graphene by the systematic attenuation of the 2D peak under hBN (Figure \ref{fig:figS2}b).

\subsection*{S3}

\begin{figure}[H]
    \centering
    \includegraphics[width=\linewidth, trim=0 5 10 0, clip]{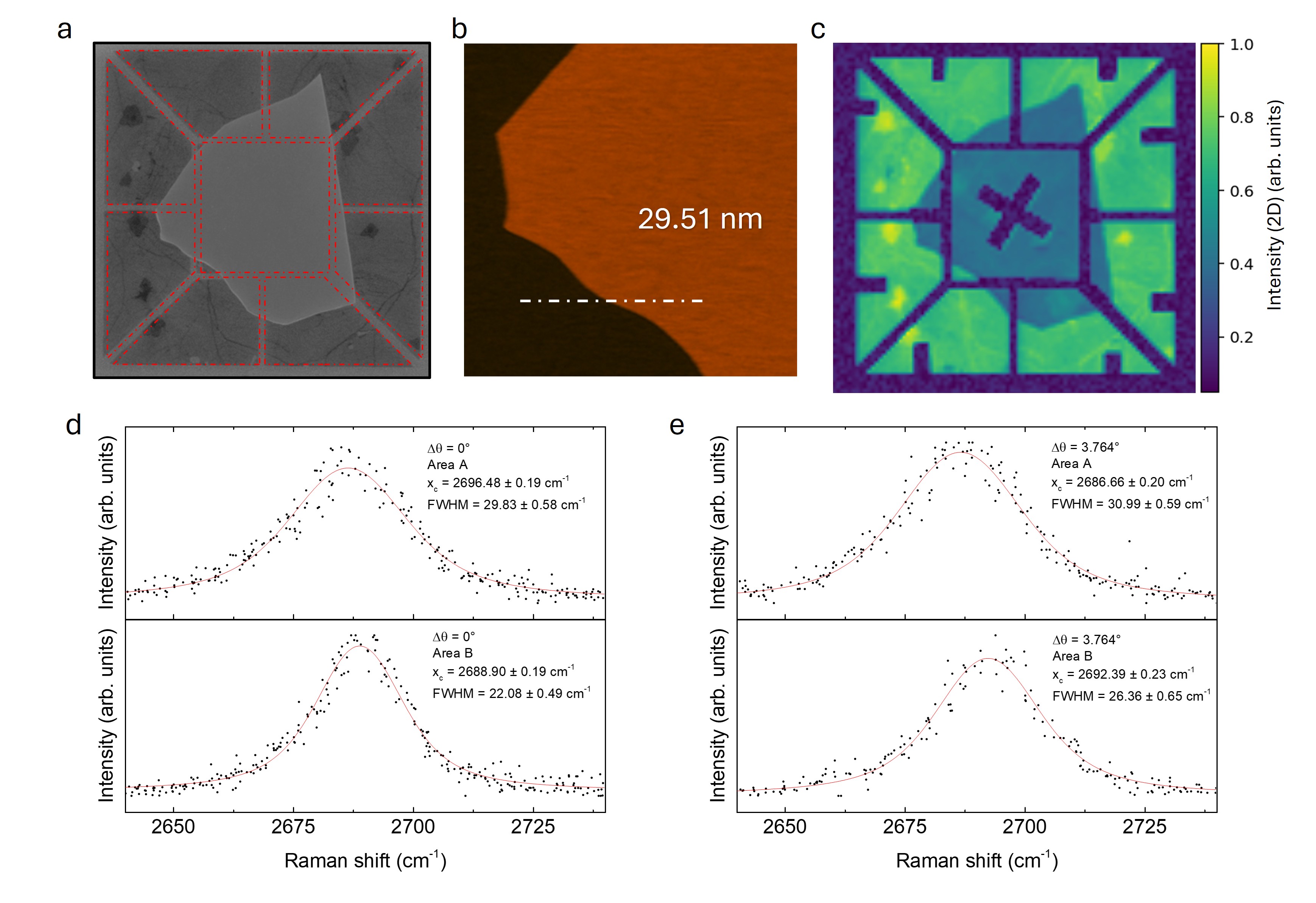}
\caption{
a) SEM image of the device. Image-based analysis using the Python routine yields a twist of 3.764°.
b) AFM topography at the hBN flake (tapping mode), showing a thickness of 29.51 nm.
c) Raman map of the graphene 2D-band intensity, normalized to the local maximum; the bare graphene area (Area A) and the graphene/hBN overlap (Area B) are clearly distinguishable.
d) Average Raman spectra of the graphene 2D band before actuation: Area A, bare graphene (top), and Area B, graphene/hBN overlap (bottom). Red curves are Voigt fits used to extract peak center and FWHM.
e) Average Raman spectra of the graphene 2D band after actuation: Area A, bare graphene (top), and Area B, graphene/hBN overlap (bottom). Red curves are Voigt fits used to extract peak center and FWHM.}
 \label{fig:figS3}
\end{figure}

SEM imaging (Figure \ref{fig:figS3}a) shows the hBN flake on the patterned graphene before electrostatic actuation. SEM image-based analysis performed with the Python routine yields a CCW twist of 3.764°. The AFM measurement at the flake edge (Figure \ref{fig:figS3}b) yields a thickness of 29.51 nm. The 2D Raman intensity map of the heterostructure (Figure \ref{fig:figS3}c) clearly distinguishes the bare graphene region from the graphene/hBN overlap, with the latter showing the expected reduction in normalized 2D intensity due to optical interference. As in Samples S1 and S2, the corresponding FWHM maps of the 2D peak before and after actuation (Figure \ref{fig:figS3}d and Figure \ref{fig:figS3}e) reveal that the bare graphene region maintains a uniform FWHM, with no significant variations between the two states, indicating that the uncovered graphene lattice remains unaffected by the actuation. In contrast, the covered area shows a measurable broadening of the 2D peak linewidth after actuation.
Average Raman spectra extracted from both regions (Figure \ref{fig:figS3}d–e) confirm this behavior. In Area A, the 2D peak position changes from $x_C = 2686.48 \pm 0.19 \ \text{cm}^{-1}$ to $x_C = 2686.66 \pm 0.20 \ \text{cm}^{-1}$, and the FWHM varies from $29.83 \pm 0.58 \ \text{cm}^{-1}$ to $30.99 \pm 0.59 \ \text{cm}^{-1}$ following actuation; both variations are within the experimental uncertainty and are not statistically significant. Conversely, in Area B the 2D peak shifts from $2688.90 \pm 0.19 \ \text{cm}^{-1}$ to $2692.39 \pm 0.23 \ \text{cm}^{-1}$, and the FWHM increases from $22.08 \pm 0.49 \ \text{cm}^{-1}$ to $26.36 \pm 0.65 \ \text{cm}^{-1}$, indicating a twist between stator and rotor.
\subsection*{S4}
\begin{figure}[H]
    \centering
    \includegraphics[width=0.9\linewidth, trim=0 5 10 0, clip]{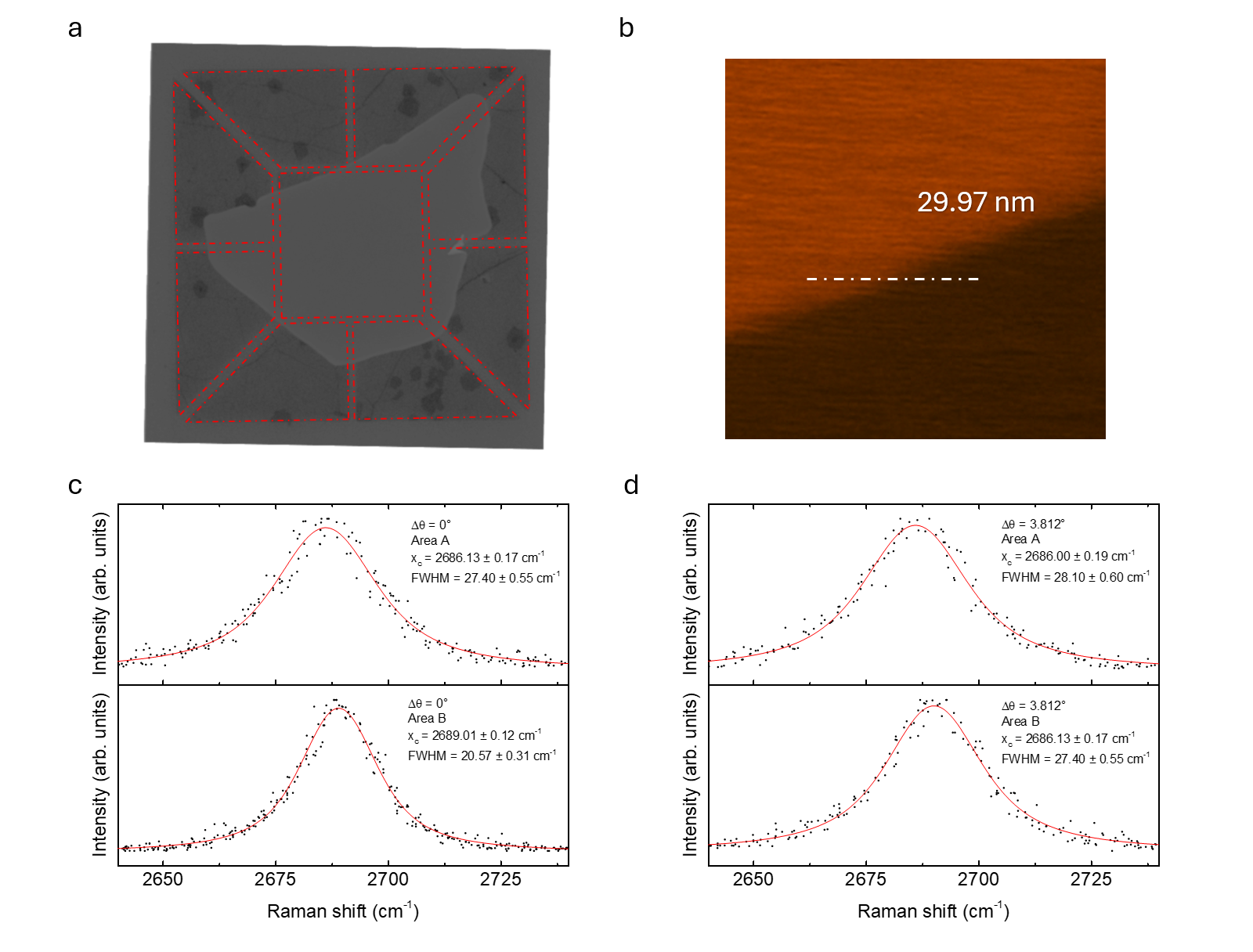}
\caption{
a) SEM image of the device. Image-based analysis using the Python routine yields a twist of 3.764°.
b) AFM topography at the hBN flake (tapping mode), showing a thickness of 29.51 nm.
c) Average Raman spectra of the graphene 2D band before actuation: Area A, bare graphene (top), and Area B, graphene/hBN overlap (bottom). Red curves are Voigt fits used to extract peak center and FWHM.
d) Average Raman spectra of the graphene 2D band after actuation: Area A, bare graphene (top), and Area B, graphene/hBN overlap (bottom). Red curves are Voigt fits used to extract peak center and FWHM.}
 \label{fig:figS4}
\end{figure}

SEM imaging (Figure \ref{fig:figS4}a) shows the hBN flake on the patterned graphene before electrostatic actuation. SEM image-based analysis performed with the Python routine yields a CW twist of 3.812°. The AFM measurement at the flake edge (Figure \ref{fig:figS4}b) yields a thickness of 29.97 nm. The FWHM of the 2D peak before and after actuation (Figure \ref{fig:figS4}c and Figure \ref{fig:figS4}d) reveal that the bare graphene region maintains a uniform FWHM, with no significant variations between the two states, indicating that the uncovered graphene lattice remains unaffected by the actuation. In contrast, the covered area shows a measurable broadening of the 2D peak linewidth after actuation. In the graphene-only area, the 2D peak position changes from $x_C = 2686.13 \pm 0.17 \ \text{cm}^{-1}$ to $x_C = 2686.00 \pm 0.19 \ \text{cm}^{-1}$, and the FWHM varies from $27.40 \pm 0.55 \ \text{cm}^{-1}$ to $28.10 \pm 0.60 \ \text{cm}^{-1}$ following actuation; both variations are within the experimental uncertainty and are not statistically significant. Conversely, in graphene/hBN overlapping area the 2D peak shifts from $2689.01 \pm 0.12 \ \text{cm}^{-1}$ to $2686.13 \pm 0.17 \ \text{cm}^{-1}$, and the FWHM increases from $20.57 \pm 0.31 \ \text{cm}^{-1}$ to $27.40 \pm 0.55 \ \text{cm}^{-1}$, indicating a twist between stator and rotor.
\subsection*{S5}
\begin{figure}[H]
    \centering
    \includegraphics[width=0.9\linewidth, trim=0 5 10 0, clip]{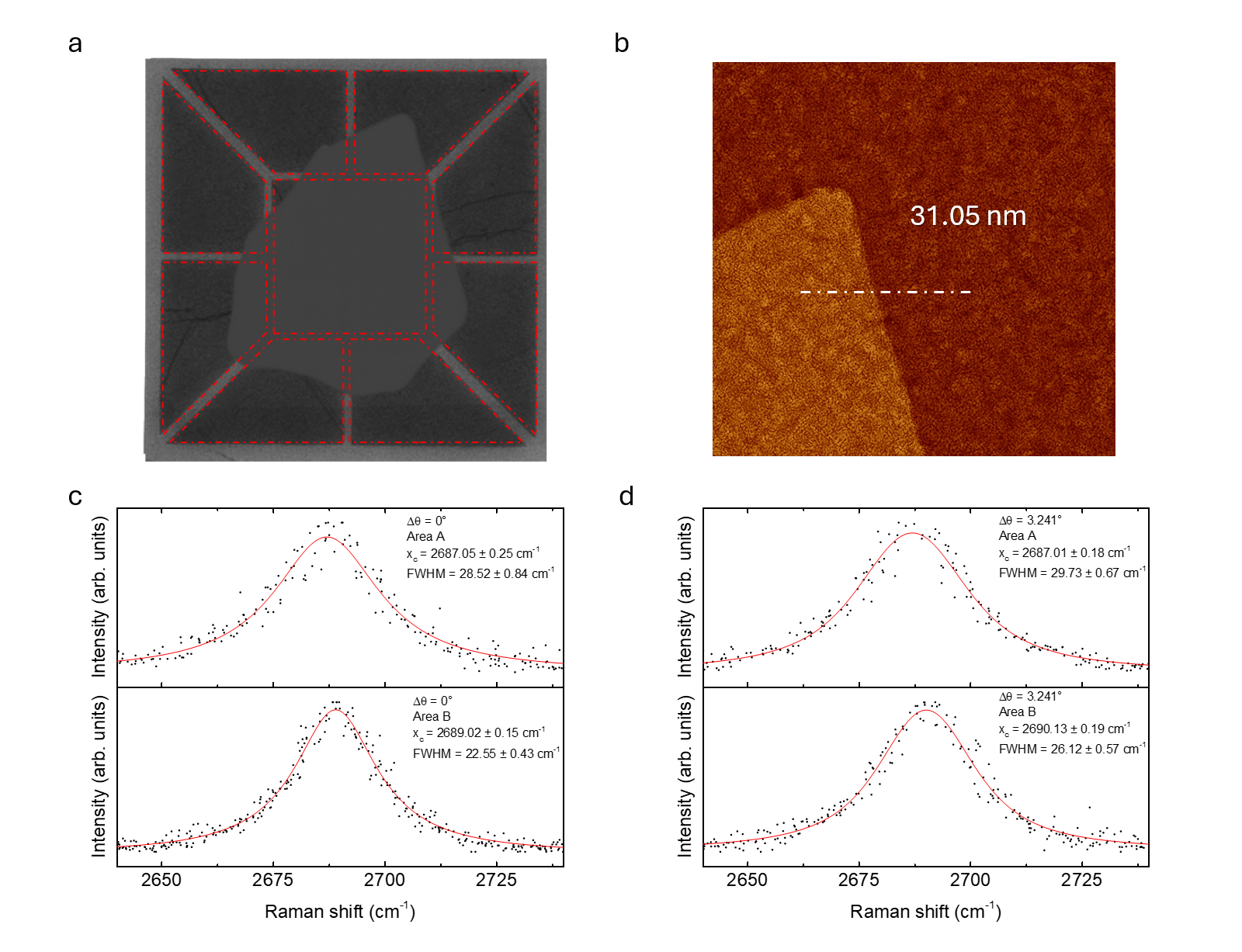}
\caption{
a) SEM image of the device. Image-based analysis using the Python routine yields a twist of 3.764°.
b) AFM topography at the hBN flake (tapping mode), showing a thickness of 29.51 nm.
c) Average Raman spectra of the graphene 2D band before actuation: Area A, bare graphene (top), and Area B, graphene/hBN overlap (bottom). Red curves are Voigt fits used to extract peak center and FWHM.
d) Average Raman spectra of the graphene 2D band after actuation: Area A, bare graphene (top), and Area B, graphene/hBN overlap (bottom). Red curves are Voigt fits used to extract peak center and FWHM.}
 \label{fig:figS5}
\end{figure}

SEM imaging (Figure \ref{fig:figS5}a) shows the hBN flake on the patterned graphene before electrostatic actuation. SEM image-based analysis performed with the Python routine yields a CCW twist of 3.241°. The AFM measurement at the flake edge (Figure \ref{fig:figS5}b) yields a thickness of 31.05 nm. The FWHM of the 2D peak before and after actuation (Figure \ref{fig:figS5}c and Figure \ref{fig:figS5}d) reveals that the bare graphene region maintains a uniform FWHM, with no significant variations between the two states, indicating that the uncovered graphene lattice remains unaffected by the actuation. In contrast, the covered area shows a measurable broadening of the 2D peak linewidth after actuation. In the graphene-only area, the 2D peak position changes from $x_C = 2687.05 \pm 0.25 \ \text{cm}^{-1}$ to $x_C = 2687.01 \pm 0.18 \ \text{cm}^{-1}$, and the FWHM varies from $28.52 \pm 0.84 \ \text{cm}^{-1}$ to $29.73 \pm 0.67 \ \text{cm}^{-1}$ following actuation; both variations are within the experimental uncertainty and are not statistically significant. Conversely, in the graphene/hBN overlapping area, the 2D peak shifts from $2689.02 \pm 0.15 \ \text{cm}^{-1}$ to $2690.13 \pm 0.19 \ \text{cm}^{-1}$, and the FWHM increases from $22.55 \pm 0.43 \ \text{cm}^{-1}$ to $26.12 \pm 0.57 \ \text{cm}^{-1}$, indicating a twist between stator and rotor.
\subsection*{S6}
\begin{figure}[H]
    \centering
    \includegraphics[width=0.9\linewidth, trim=0 5 10 0, clip]{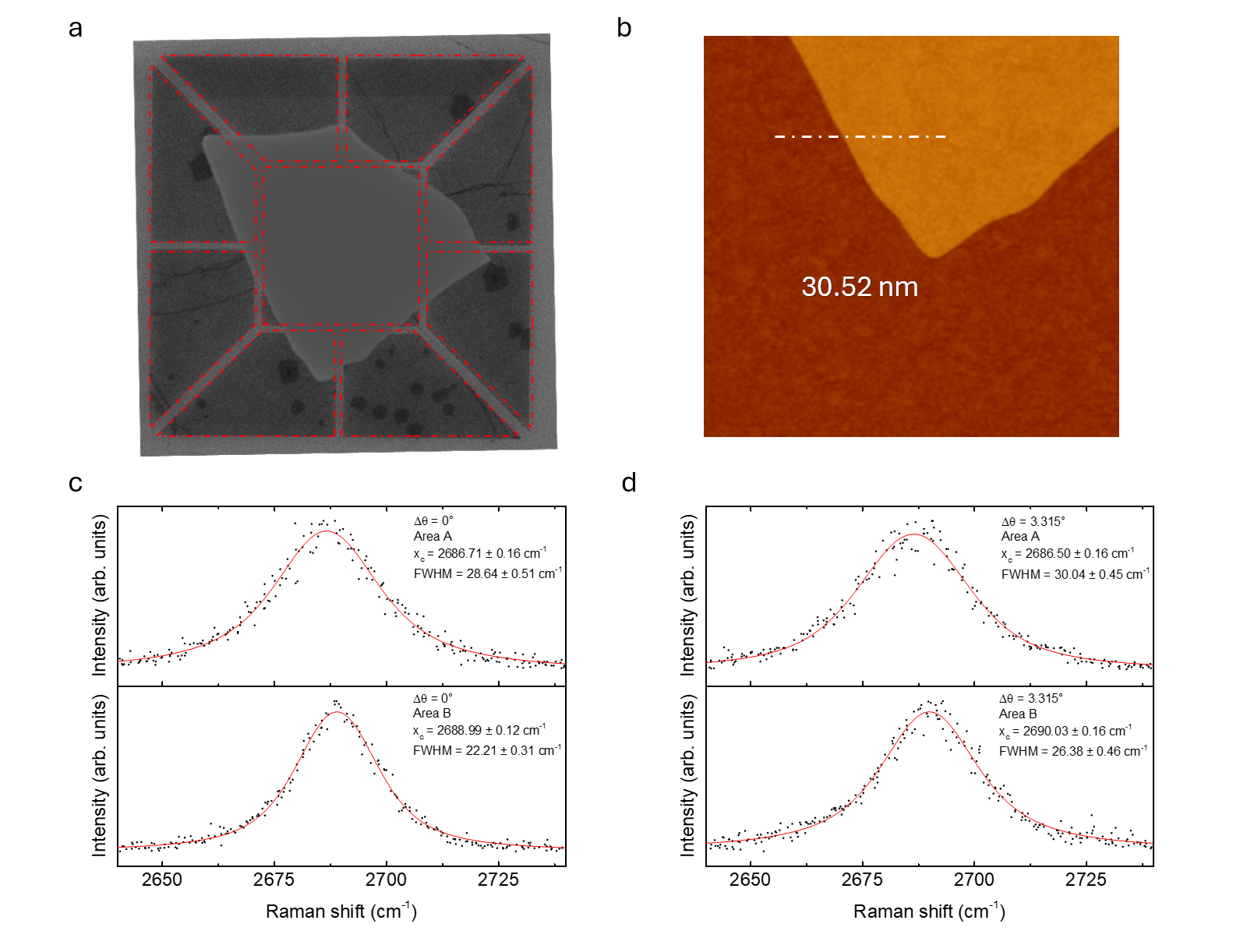}
\caption{
a) SEM image of the device. Image-based analysis using the Python routine yields a twist of 3.764°.
b) AFM topography at the hBN flake (tapping mode), showing a thickness of 29.51 nm.
c) Average Raman spectra of the graphene 2D band before actuation: Area A, bare graphene (top), and Area B, graphene/hBN overlap (bottom). Red curves are Voigt fits used to extract peak center and FWHM.
d) Average Raman spectra of the graphene 2D band after actuation: Area A, bare graphene (top), and Area B, graphene/hBN overlap (bottom). Red curves are Voigt fits used to extract peak center and FWHM.}
 \label{fig:figS6}
\end{figure}

SEM imaging (Figure \ref{fig:figS6}a) shows the hBN flake on the patterned graphene before electrostatic actuation. SEM image-based analysis performed with the Python routine yields a CCW twist of 3.315°. The AFM measurement at the flake edge (Figure \ref{fig:figS6}b) yields a thickness of 29.97 nm. The FWHM of the 2D peak before and after actuation (Figure \ref{fig:figS6}c and Figure \ref{fig:figS6}d) reveals that the bare graphene region maintains a uniform FWHM, with no significant variations between the two states, indicating that the uncovered graphene lattice remains unaffected by the actuation. In contrast, the covered area shows a measurable broadening of the 2D peak linewidth after actuation. In the graphene-only area, the 2D peak position changes from $x_C = 2686.71 \pm 0.16 \ \text{cm}^{-1}$ to $x_C = 2686.50 \pm 0.16 \ \text{cm}^{-1}$, and the FWHM varies from $28.64 \pm 0.51 \ \text{cm}^{-1}$ to $30.04 \pm 0.45 \ \text{cm}^{-1}$ following actuation; both variations are within the experimental uncertainty and are not statistically significant. Conversely, in the graphene/hBN overlapping area, the 2D peak shifts from $2688.99 \pm 0.12 \ \text{cm}^{-1}$ to $2690.03 \pm 0.16 \ \text{cm}^{-1}$, and the FWHM increases from $22.21 \pm 0.31 \ \text{cm}^{-1}$ to $26.38 \pm 0.46 \ \text{cm}^{-1}$, indicating a twist between stator and rotor.

\nolinenumbers

\bibliographystyle{sn-nature}   
\bibliography{biblio}

\end{document}